\documentclass[a4paper,fleqn,usenatbib,referee,epsfig,color]{mnras}

\usepackage[T1]{fontenc}
\usepackage{ae,aecompl}

\usepackage{graphicx}	
\usepackage{amsmath}	
\usepackage{amssymb}	

\newif\ifAMStwofonts
\newcommand{\be}{\begin{equation}}
\newcommand{\ee}{\end{equation}}
\newcommand{\bea}{\begin{eqnarray}}
\newcommand{\eea}{\end{eqnarray}}

\title[Magnetized galactic halos and velocity lags]{\bf Magnetized galactic halos and velocity lags }

\author[R.N. Henriksen \& J.A. Irwin]
{R. N. Henriksen$^1$\thanks{henriksn@astro.queensu.ca} and J. A. Irwin$^1$\thanks{irwin@astro.queensu.ca} \\ 
$^1$Dept. of Physics, Engineering Physics \& Astronomy, Queen's University, Kingston, Ontario, K7L 3N6, Canada\\
}

\date{Accepted XXX. Received YYY; in original form ZZZ}

\pubyear{2015}

\begin{document}
\label{firstpage}
\pagerange{\pageref{firstpage}--\pageref{lastpage}}
\maketitle

\begin{abstract}
 We present an analytic model of a magnetized galactic halo surrounding a Mestel gravitating disc. The magnetic field is taken to be in energy equipartition with the pressure dominant rotating halo gas ({\it not} with the cosmic rays) , and the whole system is in a steady state. A more flexible `anisotropic equipartition' model is also explored. A definite pressure law is required to maintain the equilibrium, but the halo density is constant. The velocity/magnetic system is scale-free. The objective is to find the rotational velocity lag in such a halo. The magnetic field is not force-free  so that angular momentum may be transported from the halo to the intergalactic medium.    We find that the `X'-shaped structure observed for halo magnetic fields can be obtained together with a simple analytic formula for the rate of decline of the velocity with height $z$.The formula also predicts the change in lag with radius, $r$.  
\end{abstract}




\section{Introduction}

For some time \citep{R2000,T2000,H2007,K2007,Ke2009,Car2010,ZR15},  there has been mounting evidence that galactic halo gas lags behind the rotation of the galactic disc. This is mainly found from observations of neutral gas in HI, but not exclusively so \citep{Car2010}. Recently a decrease in the lag with radius has also been detected \citep{ZRW15,ZR15}.   Earlier work \citep[][and references therein]{Sof1992} found a similar phenomenon in molecular gas above the disc of M82. However these authors conclude that this lag is best explained by violent ejection of the gas from the starburst nucleus, and thus is a different phenomenon from that discussed here. 

Many models have been suggested to explain this phenomenon, for example, see \citet{M2011} for references, and \citet{B2012}. However until now the organized magnetic field that is known to be present in the halos \citep{WI2015,Kr2015} has not been included. If such a field is connected to the intergalactic medium and has sufficient strength, then it is able to transport angular momentum away from the halo gas to intergalactic space.

The full magnetohydrodynamic equations describing such a coupling are complicated numerically and analytically.
In this note we simplify these equations by {\it assuming equipartition between the energy of the magnetic field and the kinetic energy of a dominant component of the halo gas} (although {\it not} with the underlying disc that is rather a `boundary condition' on the halo).

Normally `equipartition' refers to energy equilibrium between the magnetic field energy and that of the cosmic rays.  However the  model that we propose here requires equipartition with the kinetic energy of at least the pressure dominant component of the gas above the disc. This kinetic energy includes the contribution of  the disc rotation velocity near the disc. 

Moreover this assumption requires that the magnetic and velocity fields be aligned.  See equation (\ref{eq:magfield}) below for a succinct definition of this `isotropic'  equipartition, but it essentially means that B and V are aligned (or anti-aligned) in each coordinate direction, $r,~ \phi$, and $z$. 

Realizing such equipartition may require that the dynamical formation or subsequent growth of the galaxy should act to amplify the initial magnetic field. Kinematic disturbances in the (co-rotating) near-disc halo (e.g. from supernovae and star formation) are likely to establish  equipartition of the magnetic field only with the cosmic rays or the corresponding turbulence \citep{Han2009}. Recently \citep{PMS2014}, a numerical study of the MHD formation of a milky-way type galaxy has in fact been conducted.

The authors of this latter study find that the field grows early in cosmological history and reaches values in the range $6-10$ $\mu G$ relatively independently of the seed magnetic field, {\it due to galactic formation gas dynamics}. The total field energy  rarely exceeds $10-20$ \% of the total kinetic energy, but these relative magnitudes seem uncertain in the halo once the disc forms. More pertinently for the equipartition assumption (\ref{eq:magfield}), these authors find that {\it a large fraction of the volume of the galaxy contains aligned magnetic and velocity fields.} This aligment extends at least up to $2$ kpc above the nascent disc, which is the observed region of the halo lag of the milky way  galaxy \citep{Car2010}. 

 Galactic disc `magnetization' due to differential rotation of a random ensemble of supernovae induced dipoles \citep{Han2009} has also been simulated. This work shows that magnetic fields of the form to be deduced here (nearly parallel in the plane and X type above the plane) do develop. However these authors do not discuss the lagging halo phenomenon. Their magnetic field growth stops when equipartition with the cosmic rays is achieved. If these are indentified with the hot X-ray halo, then this condition may suffice for our purposes (see below). A similar structure (their model Dd) has also been found to display polarization structure similar to the observations by \citet{FT2014}.

A relevant study of the magnetic fields of the Milky Way has been summarized in spectacular visual format in figure 1 of \citet{Gfarr2015} for their JF12 field model. The field structure  near the disc and especially in the halo is much as we envisage in this note. In particular the structure helically wound on cones that extends into the intergalactic space is just what is required here. Moreover the structure projects in 2d into the observed `X' structure. One can not attest to equipartition with a dominant halo component with these observations, but the winding does suggest dynamic coupling between the field and the halo gas.
     
 The energy balance implied by equipartition is numerically feasible. Anisotropic magnetic field near the disc of say $10$ $\mu G$ together with a total gas density of $0.01$ $H~ cm^{-3}$ yields an Alfv\'en speed of $\sim 220$ $km s^{-1}$. The Alfv\'en speed ought to be comparable to the rotation speed of the disc/near-halo, as this value is, in order to transport  angular momentum from the disc to the halo gas in a steady state. The limited ability of this transport to oppose the extra-galactic drag, is the mechanism for the halo lag in the isotropic equipartition model (section \ref{sect:iso}) . 

If it is only the hot component of the halo that is in dynamical equipartion with the field, then we require it to be pressure dominant. Such equipartition would be  compatible with a smaller magnetic field (say $5$ $\mu G$) and a density of $0.003$ $cm^{-3},$ characteristic of the hot halo component near the disc. If the temperature is $\ge 10^6$ K and the gas is mainly ionized hydrogen, then the pressure of the gas is $\sim B^2/(8\pi)$  with $B\sim 5$ $\mu G$.  The rotation of the hot gas will be sub-sonic if the actual temperature is as high as $2\times 10^6$ K, and the pressure balance continues to hold very nearly. The interaction of this lagging component with  HI clouds would then be much as described in \citet{M2011}. In the event that the hot halo gas is pressure dominant, we simply assume here  that its lag is communicated to all halo components.   

There is evidence for an extended hot halo of the Milky Way \citep{Gup2012} of this kind, and indeed for a variety of galaxies \citep{LiW2013}. This includes the galaxy NGC 891 that we use below as an example of a well-studied lagging halo. It is possible that this gas extends into the intergalactic medium (IGM) with the magnetic field, and helps to anchor the field there. Indeed in quiescent galaxies it may have been accreted from the IGM and subsequently shock-heated \citep{LiW2013}.

The prize for investing in a maximally simple dynamical assumption is an explicit, parameter free, formula for the halo lag in both vertical and radial directions. The parameters reappear ultimately in the pressure gradient that is necessary to balance gravitational and  magnetic forces in the average steady state. Magnetic tension balances the inertial forces due to our  assumption of dynamical equipartition. A disadvantage of the assumed steady state is that causality is difficult to establish unambiguously. We do not know how it came to be the way it is assumed to be.

In this latter connection we note that {{\it It is the boundary conditions that impose disc rotation at the plane and zero rotation at large distances, which establish our predicted lag.} The large distance condition is somewhat arbitrary. However we know that at least in the case of NGC 5775, the decline to systemic velocity at large $z$ has been clearly observed \citep[e.g.][]{R2000, T2000}, and we can not avoid halo corotation at the disc.  It is the pressure gradient that is required for the average equilibrium lag to hold that must be physically realistic. 

 We explore also a variant of dynamical equipartition (the energy balance is the same) in which the poloidal magnetic and velocity fields are aligned (and equal in suitable Units) while the toroidal fields are anti-aligned (but equal in magnitude in suitable Units) (section \ref{sect:aniso}). We refer to this form as `anisotropic' equipartition as opposed to `isotropic' (same aligment in all directions) equipartition for the usual case. The anisotropic configuration follows from a slightly more physically motivated set of assumptions, but leads to a very similar lagging behaviour. The possibility encourages the following nuance in the physical interpretation of the mechanism for the halo lag. 

In the isotropic equipartition case, positive magnetic coupling (the magnetic stress, $B_\phi B_z$ in cylindrical polar coordinates, has positive sign) to the galactic disc leads to a gradual spin-down if disc gas moves up (and possibly out, following the magnetic field). The magnetic transfer of angular momentum from the disc then modulates the descent to systemic velocity at infinity, as imposed by the boundary condition. In this sense the inter-galactic medium is ultimately the source of the drag on the halo gas. Should $B_z<0$, implying $v_z<0$, accreting material is being rotationally accelerated by the disc against drag imposed by the inter-galactic medium. Any effect on the disc itself is avoided by maintaining the fixed boundary condition $v_\phi=V$ at the disc. This behaviour is very similar to that of Hartmann flow \citep[e.g.][]{LLelec} when one boundary tends to infinity. 

  In the anisotropic case, negative magnetic coupling to the intergalactic medium at infinity (drag) gradually slows down the halo gas as it rises ($B_\phi<0$ $B_z>0$) from the disc. Should the signs be reversed ($B_\phi>0$, $B_z<0$) and the gas still rising, it is the poloidal velocity and magnetic fields that are anti-aligned. We have not discussed this possibility explicitly here, but the symmetry of a global, dipole-like, galactic magnetic field \citep{Gfarr2015} implies its existence on at least one side of the galactic disc. This assumes that the gas is ejected from the disc on each side. 
Finally if the gas is accreting with $B_z<0$ and $B_\phi>0$, we return to the isotropic limit because $v_\phi>0$ by definition of the axes. 
We will find it difficult to distinguish observationally the form of the lag established by these different arrangements. 
 
We proceed to summarize the structure of this paper.  
In section \ref{sect:basicequations} we describe the basic magnetohydrodynamic equations under the equipartition assumption, together with the gravitational field of the underlying Mestel disc. Following this in section \ref{sect:iso} we find an analytic scale-free solution of these equations, which includes the gravitational field. In particular the velocities of the halo gas and the pressure law necessary for equilibrium are given. Section \ref{sect:aniso} investigates the implications of the anisotropic equipartition, which involves the viscosity explicitly. We find that there is a regime of effective overlap with the simpler isotropic model, and continue subsequently in this paper with the simpler case. In section \ref{sect:lags} we isolate the relevant formulae for the vertical and radial lagging velocities. Section \ref{sect:fields} provides figures that represent the spatial structure of the velocity/ magnetic fields. These images are useful for making qualitative comparison with observations. Section \ref{sect:test} compares our prediction for the vertical rotational velocity (and lag-rate) to relevant observations of the edge-on spiral galaxy NGC 891. A final section \ref{sect:conclusions} restates our conclusions and re-emphasizes our physical assumptions.  
\section{Model Equations: Isotropic Equipartition}
\label{sect:basicequations}
The isotropic equipartition assumption (equivalent in magnitude to the magnetic Alfv\'en number being equal to one) takes the form 
\be
{\bf B}=\sqrt{4\pi\rho}{\bf v},\label{eq:magfield}
\ee
where $\rho$ is the (constant) gas density above the disc, ${\bf v}$ is the vector velocity relative to the centre of the galaxy, and ${\bf B}$ is the vector magnetic field. We work in Gaussian electro-magnetic units. We will find therefore the velocity field and the magnetic field together, once the halo density is chosen. The disc is not a continuous extention of the halo, but rather appears as a razor thin discontinuity with its own surface mass density. The equations are similar when a minus sign is taken in equation (\ref{eq:magfield}), but we do not consider that case further in this paper, except (partially) in what we call the anisotropic case in section \ref{sect:aniso}.

We emphasize that, for gas rising from the disc,  the magnetic stress component $B_zB_\phi/4\pi >0$ in the halo, whatever the sign in equation (\ref{eq:magfield}). This implies that angular momentum is transferred from the galactic disc to the halo gas as it  ejects from the disc. The lag calculated  under this assumption is thus a consequence of magnetic coupling to the rotating disc, given the inter-galactic drag imposed by the systemic boundary condition. For gas accreting onto the disc ($B_\phi B_z<0$), drag from the intergalactic medium is slowing its approach to the disc velocity.  Either the specific angular momentum of the ejected gas is partially maintained above the systemic value starting from the disc value, or it is gradually increased from the systemic value to the disc value for infalling gas.

Proceeding with assumption (\ref{eq:magfield}) the Navier-Stokes equation reduces to (we do not apply the steady state assumption immediately)
\be
-\nabla(\frac{p}{\rho}+\frac{{\bf B}^2}{8\pi\rho})+\nu\nabla^2({\bf v})+{\bf g}=\partial_t{\bf v}.\label{eq:N-S}
\ee
Here, the notations are standard except that ${\bf g}$ is the gravitational field due to disc and collisionless halo, and $\nu$ is the kinematic viscosity. 

In addition, because of the incompressibility assumption for the halo gas, we must have 
\be
\nabla\cdot {\bf v}=\nabla\cdot {\bf B}=0.\label{eq:divergence}
\ee
The combined Amp\`ere, Faraday and Ohm equation becomes under isotropic equipartition 
\be
\partial_t{\bf B}=\eta\nabla^2{\bf B},\label{eq:fullMHD}
\ee
where the resistivity is $\eta=c^2/(4\pi\sigma)$ and $\sigma$ is the conductivity of the halo gas. Evidently, in an equipartition steady state, this reduces to  (provided $\eta\ne 0$)  
\be
\nabla^2{\bf B}=\nabla^2{\bf v}={\bf 0}. \label{eq:vecLaplace}
\ee

Equation (\ref{eq:vecLaplace}) seems to render the resistivity irrelevant. However there is an argument, found by comparing the  equation (\ref{eq:fullMHD}) to the equation for the vorticity ($\mbox{\boldmath$\omega$}\equiv \nabla\times {\bf v}$)  in an incompressible fluid (found from the curl of equation \ref{eq:N-S}) which suggests that \citep[e.g.][]{LLelec}    
\be
{\bf B}\propto \mbox{\boldmath$\omega$} \label{eq:Bomega}
\ee
iff $\nu=\eta$. 

In fact the  {\it time-dependent} equipartition equations are only self-consistent when this  equality of the diffusion coefficients holds, because the curl of equation (\ref{eq:fullMHD}) and that of equation (\ref{eq:N-S}) would otherwise not agree. Equation (\ref{eq:vecLaplace})  avoids the inconsistency in the steady state, but the relation (\ref{eq:Bomega}) might still apply (the vorticity and the magnetic field both satisfy the vector Laplace equation). We must avoid the relation (\ref{eq:Bomega}) because then we would have $\nabla\wedge {\bf v}\propto {\bf v}$, which implies a similar relation for ${\bf B}$. This would be Beltrami flow for the velocity field and the force-free condition for the magnetic field.

 Because we wish the magnetic field to exert a torque on the halo, the force-free condition must be avoided. Thus we insist that $\nu>\eta$, which inequality  aids  turbulent preservation of the magnetic field \citep{LLelec}. This is a kind of `phantom' assumption as the viscosity does not appear subsequently  in isotropic equipartition, although it does in the anisotropic case. Even in the event that $\nu=\eta$, Beltrami or force-free flow is not the only possible conclusion depending on boundary conditions, but we will assume that $\nu\ne\eta$.  

The disc of the galaxy will be taken as a `razor thin' Mestel disc rotating with uniform velocity $V\widehat{\bf e}_\phi$. Such a disc has a scale-free surface density in terms of its cylindrical radius $r$
\be
\sigma(r)=\frac{\Sigma}{\delta r},\label{eq:surfdens}
\ee
Where $\Sigma$ is a constant and $\delta$ is a convenient reciprocal scale that is arbitrary. We will introduce this more precisely below. The gravitational potential of the Mestel disc is given by 
\be
\Phi_d=\frac{2\pi G\Sigma}{\delta}sinh^{-1}(\xi)+\frac{2\pi G\Sigma}{\delta}\ln{(\delta r)},\label{eq:Mpot}
\ee
where $G$ is Newton's constant and $\xi\equiv z/r$ in cylindrical coordinates $\{r,\phi,z\}$ relative to the polar axis of the galaxy. The constant rotational velocity is related to the constant $\Sigma$ through 
\be
V^2=\frac{2\pi G \Sigma}{\delta}.\label{eq:Mrotvel}
\ee
The reciprocal scale $\delta$ may be taken so that $\delta r_o=1$ at some fiducial radius $r_o$, if this radius is a convenient disc radius. If the density is also known at $r_o$, then $\Sigma$ is fixed by equation (\ref{eq:surfdens}). Should $r_o$ be small compared to the observable disc, as for example if it is made to represent a turbulent, viscous scale, then $\delta r>>1$, but a measurement of the disc density at $r$ would still determine $\Sigma$, given $\delta$.

The Mestel disc is actually an example of an isothermal or homothetic self-similarity in terms of the invariant $\xi$ \citep[e.g.][]{Hen2012,Hen2015}, because the existence of the multiplicative constant $V^2$ renders a Dimensionless logarithmic potential compatible with isothermal/homothetic Self-Similarity.  Because of this latter freedom a spherically symmetric, gravitating, dark or stellar halo may be added to the source of the halo gravitational field in the form (subscript c for `collisionless')
\be
\Phi_c=\lambda V^2 \ln{(\delta r\sqrt{1+\xi^2}}),\label{eq:darkpot}
\ee
where $\lambda$ is an Dimensionless constant, which gives the relative importance of the disc and collisionless halo gravitational fields.
  
The gravitational field due to the disc is found from ${\bf g}_d=-\nabla\Phi_d$, which yields 
\be
{\bf g}_d=\frac{V^2}{r}\left(\frac{\xi}{\sqrt{1+\xi^2}}-1\right)\widehat{\bf e}_r-V^2\frac{1}{r\sqrt{1+\xi^2}}~\widehat{\bf e}_z.\label{eq:discfield}
\ee
The field due to the collisionless spherical halo is ${\bf g}_c=-\nabla\Phi_c$ or
\be
{\bf g}_c=-\frac{V^2}{r}\frac{\lambda}{1+\xi^2}\widehat{\bf e}_r-\frac{V^2}{r}\frac{\lambda\xi}{1+\xi^2}\widehat{\bf e}_z.\label{eq:halofield}
\ee
We assume equilibrium of the halo gas under the sum of these two fields, namely ${\bf g}$, plus magnetic stresses.

\section{ Scale-Free solution}\label{sect:iso}

In the steady state equation (\ref{eq:vecLaplace}) applies, so that in cylindrical coordinates with axial symmetry  and a steady state,  equation (\ref{eq:N-S}) yields 
\bea
-\partial_r(\frac{p}{\rho})-\partial_r(\frac{{\bf v}^2}{2})+g_r&=&0,\nonumber\\
-\partial_z(\frac{p}{\rho})-\partial_z(\frac{{\bf v}^2}{2})+g_z&=&0,\label{eq:coordN-S}
\eea
where ${\bf g}={\bf g}_d+{\bf g}_c$ and the disc and collisionless halo components are given in the previous section. In order to find a scale-free solution it suffices to require 
\be
{\bf v}=V{\bf w}(\xi),\label{eq:scaling}
\ee
which implies by equation (\ref{eq:magfield}) a similar form for ${\bf B}$. Proceeding with this assumption in equations (\ref{eq:coordN-S}), remembering the form of $\xi$, and then adding the equations yields
\be
r\partial_r(\frac{p}{\rho})+z\partial_z(\frac{p}{\rho})=-V^2(\lambda+1),\label{eq:sound}
\ee
which has the general solution in terms of an arbitrary  function $P(\xi)$ 
\be
\frac{p}{\rho}=-V^2(\lambda+1)\ln{\delta r}+V^2P(\xi).\label{eq:solsound}
\ee 

If we substitute this last expression into either one of equations (\ref{eq:coordN-S}) we obtain an expression for the pressure function in terms of the velocity as 
\be
\frac{d}{d\xi}(P(\xi)+\frac{{\bf w}^2}{2})=-\frac{1}{\sqrt{1+\xi^2}}-\frac{\lambda\xi}{1+\xi^2}.\label{eq:hydrostatic}
\ee 
To complete the solution we must find the velocity field compatible with equations (\ref{eq:divergence}) and (\ref{eq:vecLaplace}).

Applying axial symmetry the divergence condition (\ref{eq:divergence}) yields, where $()'\equiv d()/d\xi$,
\be
 \xi w_r'-w_r-w_z'=0.\label{eq:expdivergence}
\ee
Equation (\ref{eq:vecLaplace}) becomes explicitly assuming axial symmetry 
\bea
(1+\xi^2)w_r''+\xi w_r'-w_r&=&0,\nonumber\\
(1+\xi^2)w_\phi''+\xi w_\phi'-w_\phi &=& 0,\label{eq:expvecLaplace}\\
(1+\xi^2)w_z''+\xi w_z'&=&0,\nonumber
\eea
These equations for the velocity are over determined, but one can find a self-consistent solution. 

The $z$ component of equations (\ref{eq:expvecLaplace}) integrates to give generally 
\be
w_z=c_1\ln{(\xi+\sqrt{1+\xi^2})}+c_2\equiv c_1sinh^{-1}(\xi)+c_2,\label{eq:wz}
\ee
so that unless $c_1=0$ there is a logarithmic divergence at large $\xi\equiv z/r$. Nevertheless $w_z$ is constant on cones $\xi=const.$, and when $\xi$ is large the cone collapses onto the axis of the galaxy. This is similar (although less violent) to the behaviour of the radial velocity found below. We can avoid this divergence by taking $c_1=0$ and $w_z=c_2$. This would describe constant mass ejection into the halo from the disc (or constant infall). Allowing the divergent behaviour at large $\xi$ would correspond to a central AGN behaviour.

The incompressible condition (\ref{eq:expdivergence}) generally  requires $\xi w_r'-w_r=w_z'$, which may be solved  for $w_r$ as  
\be
w_r=c_3~\xi-c_1\sqrt{1+\xi^2},\label{eq:wr}
\ee 
where $c_3$ is a new constant.
This solution also satisfies the radial component of equations (\ref{eq:expvecLaplace}). 
The strong divergence at large $\xi$ appears unacceptable, so that we might set $c_3=c_1$. However if we wish the Mestel disc at $\xi=0$ to be in equilibrium with $w_r=0$, we must set $c_1=0$. This renders $w_z=c_2$ constant and $w_r=c_3\xi$, which diverges at large $\xi$. We proceed with this special case in all of our figures.

The unique  dependence on $\xi$ of $w_r$ implies that the radial velocity takes finite values on cones centered on the nucleus of the galaxy and wrapped around its axis. The velocity is infinite  on the galactic axis and it is zero on the galactic disc. Despite the infinite radial velocity on the axis, the radial mass flux is finite there.  
Near the disc the presence  of $v_r$ allows the total velocity field and  magnetic field to take on an `X-shape' above the plane, which is desirable \citep{Kr2009,Kr2015}. We will therefore  proceed with $c_3\ne 0$, but we do not expect to apply this solution to the axis of the galaxy. In fact, our figures show that the region of interest above the disc  never approaches the axial region.

The azimuthal velocity component of equations (\ref{eq:expvecLaplace}) is of most interest to us as it describes any velocity lag in the halo. Its form depends only indirectly on the pressure and gravity fields through equation (\ref{eq:hydrostatic}). By constraining the solution for the halo rotation to show a lag, we are constraining the pressure field through the function $P(\xi)$, given the gravitation of the Mestel disc and of the collisionless halo. 

The formal solution of the azimuthal member of equations (\ref{eq:expvecLaplace}) is 
\be
w_\phi=c_4~\xi+c_5\sqrt{1+\xi^2}.
\ee  
We have the boundary condition at the disc as $w_\phi(0)=1$, and hence $c_5=1$. In order that the velocity vanish as $\xi\rightarrow \infty$ we must choose $c_4=-1$. This is the `systemic' boundary condition, which implies a strong frictional interaction with the intergalactic medium. With these boundary conditions  the constraint solution for the halo lag in isotropic equipartition is 
\be
w_\phi=\sqrt{1+\xi^2}-\xi,\label{eq:wphi}
\ee
which is parameter free, but assumes the necessary equilibrium. We observe that this form also goes to $1$ ($v_\phi\rightarrow V$) at a fixed height above the disc (fixed $z$) at large radius where $\xi\rightarrow 0$. This means that the galactic disc is surrounded by a cylinder rotating with the velocity $V$! This represents an asymptotic Taylor-Proudman column \citep[e.g.][]{G1980}. It leads to an observable increase in the halo rotation with radius, which we shall discuss again in an observational context. We have thus determined both the velocity and magnetic fields in the lagging halo in (isotropic) equipartition.


The pressure function $P(\xi)$ follows now from equation (\ref{eq:hydrostatic}) in the form 
\be
P(\xi)=c_6-(\frac{w_r^2+w_z^2+w_\phi^2}{2})-\frac{\lambda}{2}\ln{(1+\xi^2)}-sinh^{-1}(\xi),\label{eq:A}
\ee
where $c_6$ is a constant to be determined from $c_6=P(0)+({\bf w}(0)^2)/2$. Here $w_\phi^2$ follows from equation (\ref{eq:wphi})
while $w_r$ and $w_z$ are given generally by equations (\ref{eq:wr}) and (\ref{eq:wz}) respectively, but the simplest solution has $c_1=0$.

The total halo pressure is given by equation (\ref{eq:solsound}) so that 
\be
P(0)=\frac{1}{V^2}(\frac{p}{\rho}\bigg|_o)(r)+(1+\lambda)\ln{\delta r}=c_6-\frac{(1+c_2^2)}{2}.\label{eq:A0}
\ee
We have already discussed the behaviour of the  radial  and vertical velocities in equation (\ref{eq:A}). The apparent logarithmic potential divergence with $\xi$ in that equation  is due to the collisionless halo that was taken infinite in extent. This term  is constant on cones just as are the radial and vertical velocities.
 
 Model parameters must exist such that the pressure from equation (\ref{eq:solsound}) is positive over some reasonable vertical extent for the gaseous halo. There is always a  limit for the physical plausibility of $w_r$ at large enough $\xi$. At any radius this occurs where a vertical cylinder of this radius intersects a cone of unreasonably small opening angle and it is usually so large as to be unphysical. More directly, the condition for zero pressure becomes by equation (\ref{eq:solsound}) $P(\xi)=(\lambda+1)\ln{\delta r}$. We proceed by eliminating $c_6$ from equation (\ref{eq:A}) using equation (\ref{eq:A0}). Subsequently we  insert $w_z=c_2$ and the forms of $w_r$ and $w_\phi$ as functions of $\xi$ into equation (\ref{eq:A}). This equation may be manipulated until the zero pressure condition is on one side of the  equation. The other side then yields a condition on $\xi_m$ where the pressure tends to zero in the form 
\be
a_{s0}^2(r)+\xi_m(\sqrt{1+\xi_m^2}-\xi_m)=\frac{c_3^2\xi_m^2}{2}+\ln{(1+\xi_m^2)^{\lambda/2}(\xi_m+\sqrt{1+\xi_m^2})}.\label{eq:xicrit}
\ee
We have set $a_{s0}^2\equiv (p/\rho)_0V^{-2}$. For the parameters of figure (\ref{fig:equispace}) (that is $\lambda=1$, $c_3=0.1$, $a_{s0}^2=4$) this formula gives $\xi_m\approx 6.05$ in agreement with the illustration. Equation (\ref{eq:xicrit}) is readily solved numerically for any particular choice of parameters. If $a_{s0}^2$ is converted to a temperature, it lies in the x-ray range. This suggests once again that we are dealing with a dominant hot halo component.

We have found in this section the complete behaviour of an equipartition, steady, axially-symmetric, model. Magnetically and kinematically the solution is scale-invariant on cones. The magnetic field and the velocity field are helical ( wound on cones when $v_r\ne 0$), becoming straight (ultimately radial) at large distances. The magnetic field is thus tightly wound at the galactic disc and anchored in the inter-galactic medium . This configuration allows angular momentum to be transferred from the halo to the  inter-galactic medium. For accreting halo material (and the positive sign in equation \ref{eq:magfield}) this gives $B_\phi B_z<0$ and the drag is explicit. For gas rising from the disc (so that $B_z>0$) $B_\phi B_z>0$ and the disc is moderating the imposed descent to the systemic velocity. However one would expect on the basis of causality (not included in the steady state) that we require $B_\phi B_z<0$ also in this configuration. This requires anisotropic equipartition that we consider in the next section.  

\section{Anisotropic Equipartition Halo}\label{sect:aniso}

We turn to the anisotropic possibility wherein the rising halo gas  is  directly dragged by the intergalactic medium. In this case angular momentum of the halo can be transferred to  inter-galactic space. We require $B_zB_\phi/4\pi<0$ everywhere, which is achieved by the anisotropic version of equipartition for rising gas. This configuration offers an intriguing magnetic link between the ejection activity of the disc and the inter-galactic medium.  The predicted lag can nevertheless be similar to that found above for isotropic equipartition.

We proceed with the same form of the scale-free solution as in the previous section. However in place of equation (\ref{eq:magfield}) we require only 
\be
\nabla\times({\bf v}\times {\bf B})={\bf 0},\label{eq:perfectMHD}
\ee 
which follows from the steady MHD equation when the resistivity is zero. The radial component of this equation gives 
\be
\frac{w_r}{b_r}=\frac{w_z}{b_z}~(=1),\label{eq:polratio}
\ee
which ratio we set equal to one to enforce poloidal equipartition (it might also be $-1$, but then $b_\phi=v_\phi$ for anisotropy). The $z$ component of this equation reduces to an identity, while the azimuthal component can be written after some manipulation as 
\be
\frac{d}{d\xi}(w_\phi-b_\phi)(w_z-\xi w_\phi)=w_r(w_\phi-b_\phi).\label{eq:MHDplusdiv}
\ee
We have used equation (\ref{eq:polratio}) to replace poloidal field components with poloidal velocity components in this relation, and subsequently used the divergence condition (\ref{eq:expdivergence}), in order to obtain this form.  

The azimuthal component of the Navier-Stokes equation no longer reduces to equation (\ref{eq:vecLaplace}). In fact it becomes 
\be
(w_z-\xi w_r)\frac{d}{d\xi}(w_\phi-b_\phi)+w_r(w_\phi-b_\phi)= \frac{\overline{\nu}\delta}{V}\left(w_\phi''(1+\xi^2)+\xi w_\phi'-w_\phi\right).\label{eq:N-Sphi}
\ee
We have extended the Self-Similar symmetry to include the viscosity according to Dimensional analysis as dictated by the constant velocity of the underlying Mestel disc. That is, the compatible kinematic viscosity is $\propto Vr$. Hence
\be
\nu=\overline{\nu}e^{\delta R}\equiv \overline{\nu}\delta r,\label{eq:SSvisc}
\ee
where $\overline\nu$ is a constant.
The general theory of such analysis has recently been discussed (e.g. \cite{Hen2015}) and is summarized in the footnote. \footnote{The Dimensional co-vectors are ${\bf d}_\nu=(-1,2,0)$ and $\alpha=\delta$ so that finally ${\bf d}_\nu=(0,1,0)$.}
If we combine equation (\ref{eq:MHDplusdiv}) with equation (\ref{eq:SSvisc}), and use the anisotropic condition 
\be
b_\phi=-w_\phi,\label{eq:anisotropic}
\ee
 we obtain the important relation
\be
4w_rw_\phi=\frac{\overline{\nu}\delta}{V}\left(w_\phi''(1+\xi^2)+\xi w_\phi'-w_\phi\right).\label{eq:newlag}
\ee
This is the  equation for the halo lag with the assumption of anisotropic equipartition. If $w_r=0$, we return to the lag of the isotropic model that is given by setting the RHS equal to zero. In that case the fields are wrapped on cylinders with no `X structure'.  In the anisotropic model we must find equations for both $w_r$ and $w_z$.

The final form of the Self-Similar symmetry is the expression for the pressure. Equation (\ref{eq:solsound}) does not follow in such a compelling way from the radial and vertical dynamical equations in this case, because the radial component has the extra inertial term $w_\phi^2-b_\phi^2$ on the RHS. We can still assume that 
\be
\frac{p}{\rho}=V^2P(\xi)-V^2(\lambda +1)\ln{\delta r},
\ee
on physical grounds because it is the most general form in isothermal Self-Similarity. The radial and vertical dynamical equations now become after the usual substitutions 
\bea
&\xi\frac{d}{d\xi}(P+\frac{w_r^2+w_z^2+w_\phi^2}{2})+(w_\phi^2-b_\phi^2)+\frac{\overline{\nu}\delta}{V}(w_r''(1+\xi^2)+\xi w_r'-w_r)+(\lambda +1)+\frac{rg_r}{V^2}=0,\label{eq:anisoDr}\\
&-\frac{d}{d\xi}(P+\frac{w_r^2+w_\phi^2+w_\phi^2}{2})+\frac{\overline{\nu}\delta}{v}(w_z''(1+\xi^2)+\xi w_z')+\frac{rg_z}{V^2}=0\label{eq:anisoDz}
\eea

To proceed we need a small lemma. Differentiating the divergence condition (\ref{eq:expdivergence}) yields $\xi w_r''=w_z''$ and this together with the divergence condition shows that 
\be
(1+\xi^2)w_r''+\xi w_r'-w_r\equiv \frac{1}{\xi}(w_z''(1+\xi^2)+\xi w_z').\label{eq:lemma}
\ee
This lemma allows us to eliminate these two expressions, first by multiplying equation (\ref{eq:anisoDr}) by $\xi$ and then subtracting equation (\ref{eq:anisoDz}) to obtain 
\be
 (1+\xi^2)\frac{d}{d\xi}(P+\frac{w_r^2+w_z^2+b_\phi^2}{2})+(\lambda+1)\xi +\sqrt{1+\xi^2}-\xi=0.\label{eq:Ddiff}
\ee
We have put in the components of the gravitational field explicitly and used again the anisotropic condition. Our actual working equation follows by using equation (\ref{eq:Ddiff}) in equation (\ref{eq:anisoDz}) to eliminate the gradient term. We use the anisotropic condition and the explicit form of $g_z$ to find  once again that in fact    
\be
(1+\xi^2)w_z''+\xi w_z'=0.
\ee
Hence by the lemma, $w_r$ and $w_z$ are given by the general solutions (\ref{eq:wr}) and (\ref{eq:wz}) respectively. We need now only to solve equation (\ref{eq:newlag}) for $w_\phi$ after which equation (\ref{eq:Ddiff}) yields the pressure function $P(\xi)$. The latter is identical to equation (\ref{eq:hydrostatic}) but for the more general solution for $w_\phi$ that follows from equation (\ref{eq:newlag}). In particular equations (\ref{eq:A}) and (\ref{eq:A0}) continue to apply. 
In figure (\ref{fig:anisoequipart}) we compare two anisotropic equipartition cases to the isotropic case.

\begin{figure}
\begin{tabular}{cc} 
\rotatebox{0}{\scalebox{.8} 
{\includegraphics{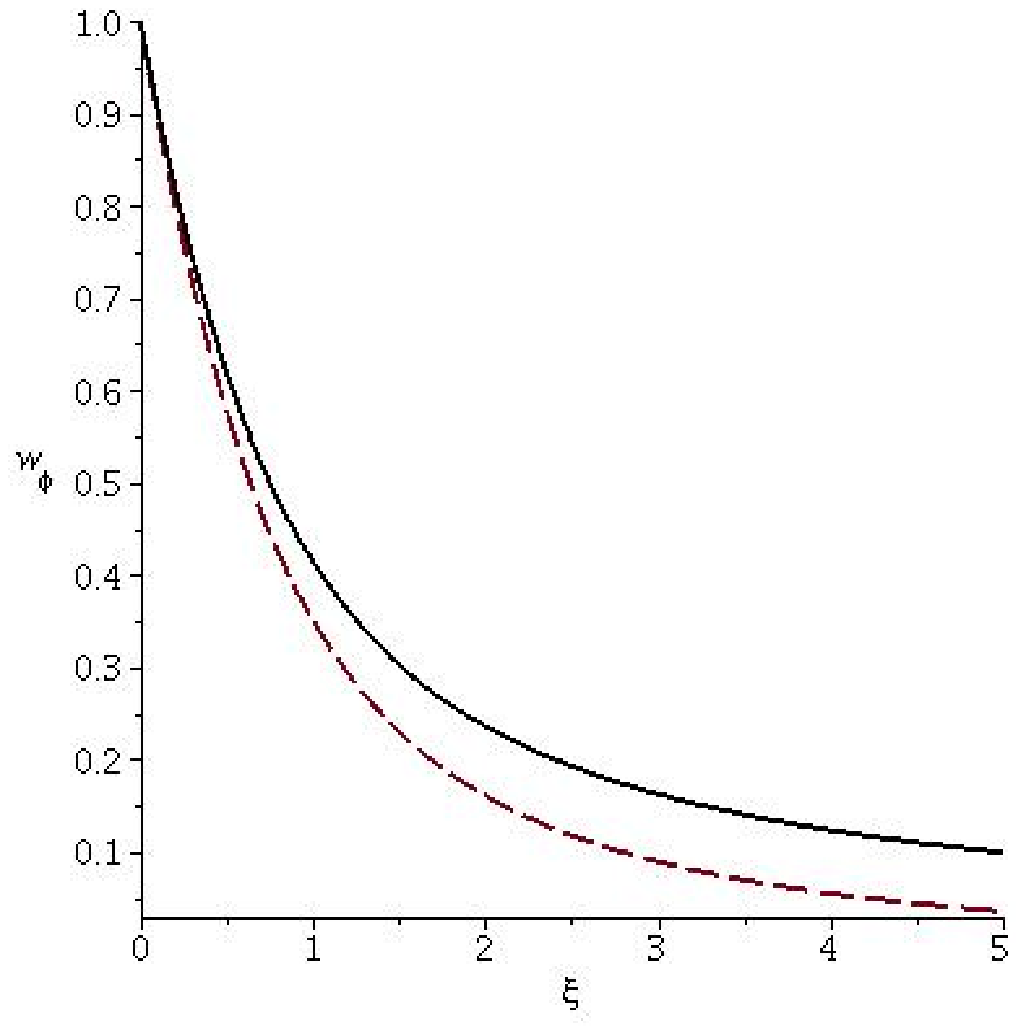}}}&
\rotatebox{0}{\scalebox{.8} 
{\includegraphics{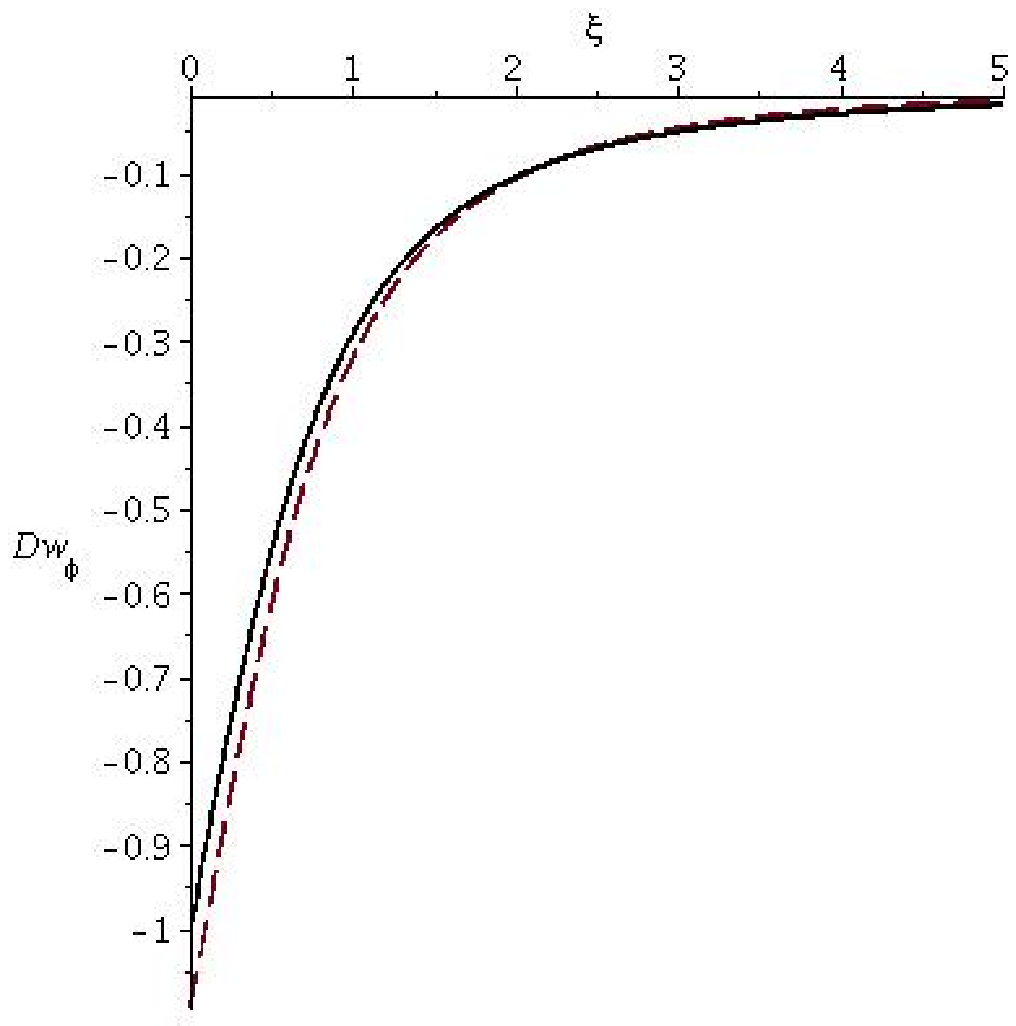}}}\\
{\rotatebox{0}{\scalebox{.8} 
{\includegraphics{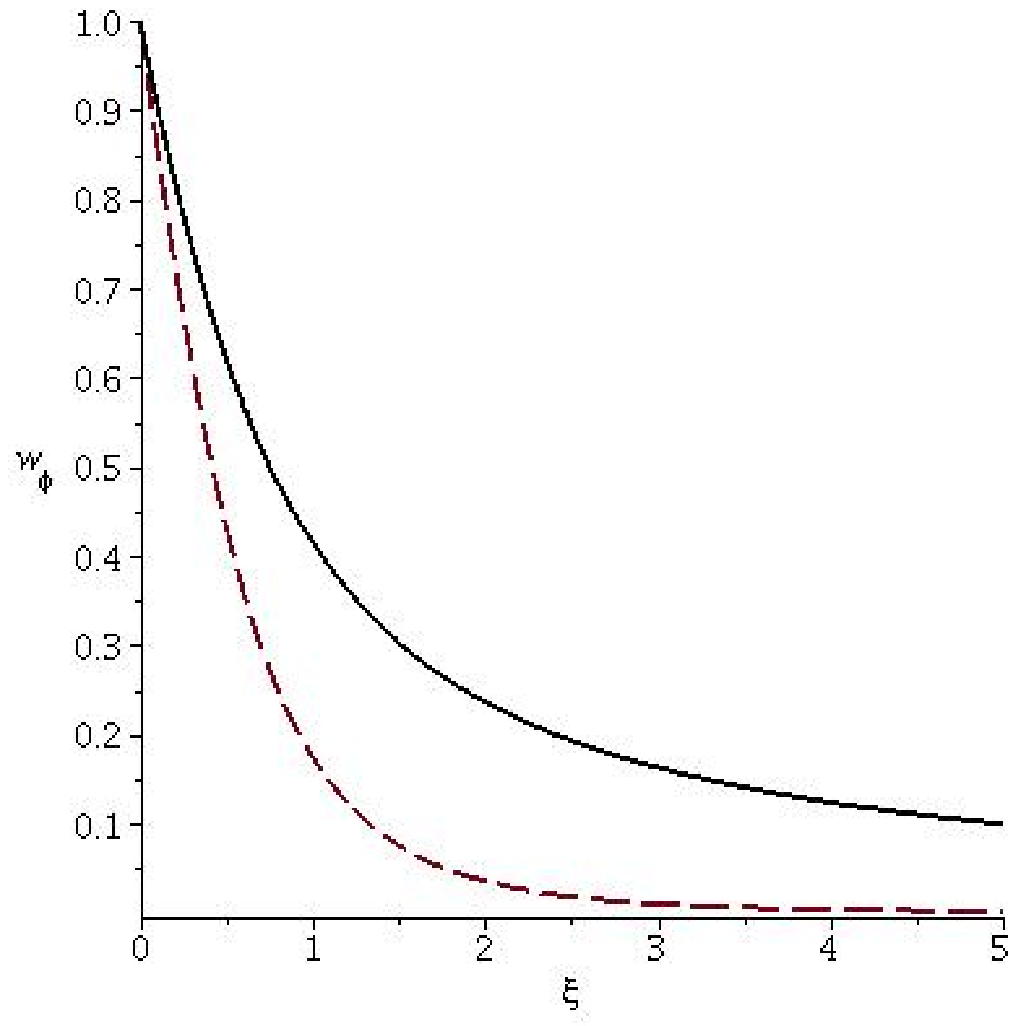}}}}&
\rotatebox{0}{\scalebox{.75} 
{\includegraphics{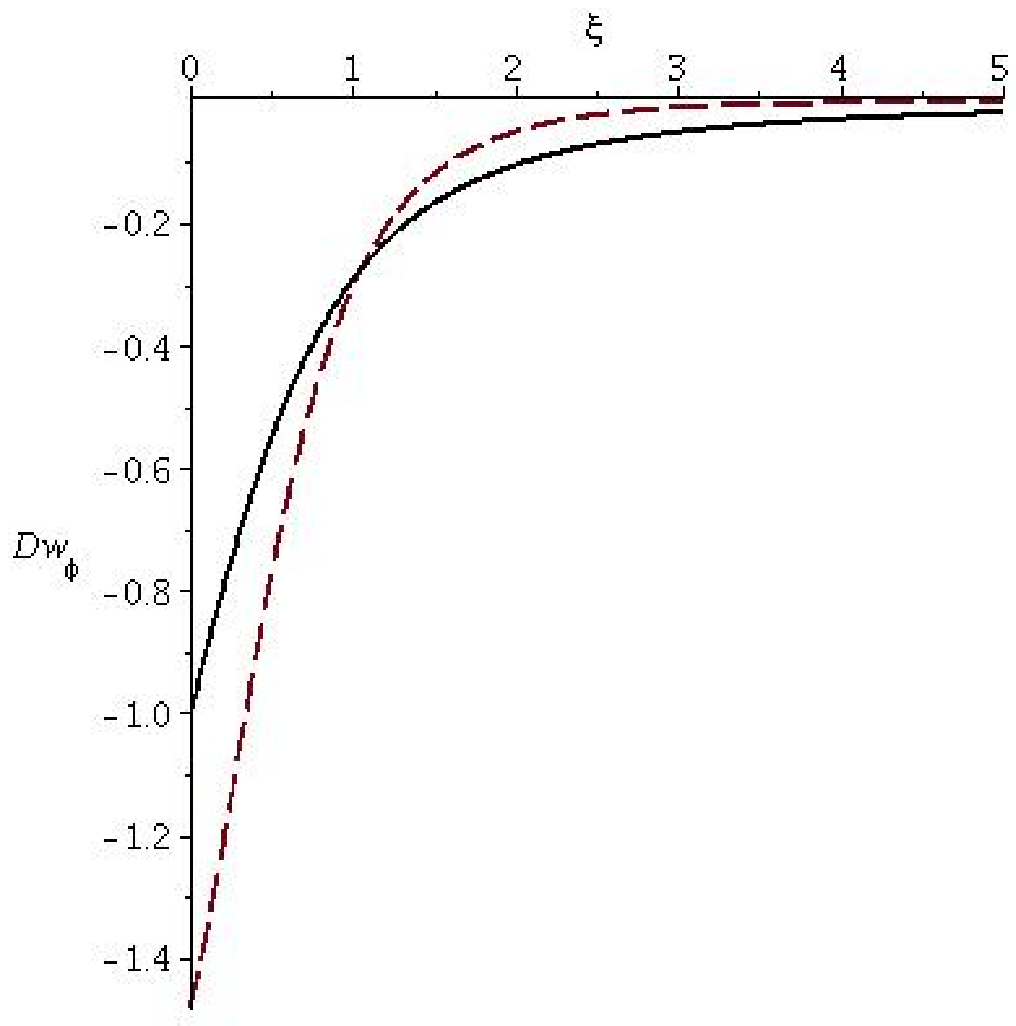}}}
\end{tabular}
\caption{At upper left we compare the anisotropic lag velocity (dashed line) to the isotropic velocity lag (solid line), both as functions of $\xi=z/r$. Here $c_1=0$, $c_3=0.1$ and $k \equiv V/(\overline{\nu}\delta)=1$. On the upper right figure we plot the actual slope in units of $\delta V$ for the anisotropic (dashed) and isotropic cases (solid). The lower half of the figure shows the same two quantities with the same constants except that $k=10$.   }    
\label{fig:anisoequipart}
\end{figure}

We recall that $\delta$ is an arbitrary reciprocal scale  and may be chosen so as to resolve the scale deemed physically important. In this way its choice permits a kind of adjustable `magnification' by which to view the system. Only $\Sigma/\delta$ plus some characteristic radius need be given to fix the physical disc surface density everywhere. The quantity $\xi$ is scale invariant so that only the ratio $\delta r$ (with $r$ a fiducial or characteristic radius) varies.

If for example we suppose $1/\delta$ to be on the scale of the turbulence creating the viscosity, then $k\equiv V/(\overline{\nu}\delta)=1$. This  increases the value of $1/\delta$ relative to the reciprocal of a characteristic disc radius. Consequently  the disc surface density  constant $\Sigma$ would be reduced proportionately in order to keep the physical surface density the same, according  to  equation (\ref{eq:surfdens})}.  This choice renders $\delta r\sim 10$, given a characteristic disc radius $r$.
Taking in addition $c_1=0$, $c_3=0$, we obtain the upper pair of curves in  figure \ref{fig:anisoequipart}. The difference with the isotropic solution is probably too small to be of observational consequence although the descent to systemic velocity is steeper. 

This trend to an increasingly rapid descent is accentuated in the lower pair of figures where $k=10$ since we have now chosen $1/\delta$ such that $\delta r=1$ for a characteristic disc radius $r$. The decline to systemic velocity is seen to be much steeper than in the isotropic case. Such lag behaviour may still be acceptable at say $r=2~kpc$, where systemic velocity is reached only at $z=6-8~~kpc$. In any case one can always reduce the difference with the isotropic lag by reducing the radial velocity. For this reason we study preferentially the isotropic case below. Nevertheless the anisotropic model best describes a halo lag due to gas rising from the galaxy, if the intuitive convective causal mechanism for the steady state magnetic field (twisting the inter-galactic field) is invoked.

\section{Halo Lag, Vertically and Radially}\label{sect:lags}
Recently \citep{ZRW15,ZR15}, a radial dependence on the vertical lag rate has been detected. We show in this section how this is contained in the expression (\ref{eq:wphi}).
 
In cylindrical coordinates  at rest in the galactic centre, equation (\ref{eq:wphi}) implies that the halo azimuthal velocity (and corresponding azimuthal magnetic field) is 
\be
v_\phi\equiv \frac{B_\phi}{\sqrt{4\pi\rho}}=\frac{V}{r}(\sqrt{r^2+z^2}-z).
\label{eq:vphiexp}
\ee
which reduces to the boundary conditions, $v_\phi=V$ at $z=0$ and $v_\phi \rightarrow 0$ as $z\rightarrow \infty$. If, for example, $V=220$ km/sec and $r=z=8$ kpc, then
  $v_\phi=0.4V=88$ km/sec.   
   
\begin{figure}
\begin{tabular}{cc} 
\rotatebox{0}{\scalebox{.4} 
{\includegraphics{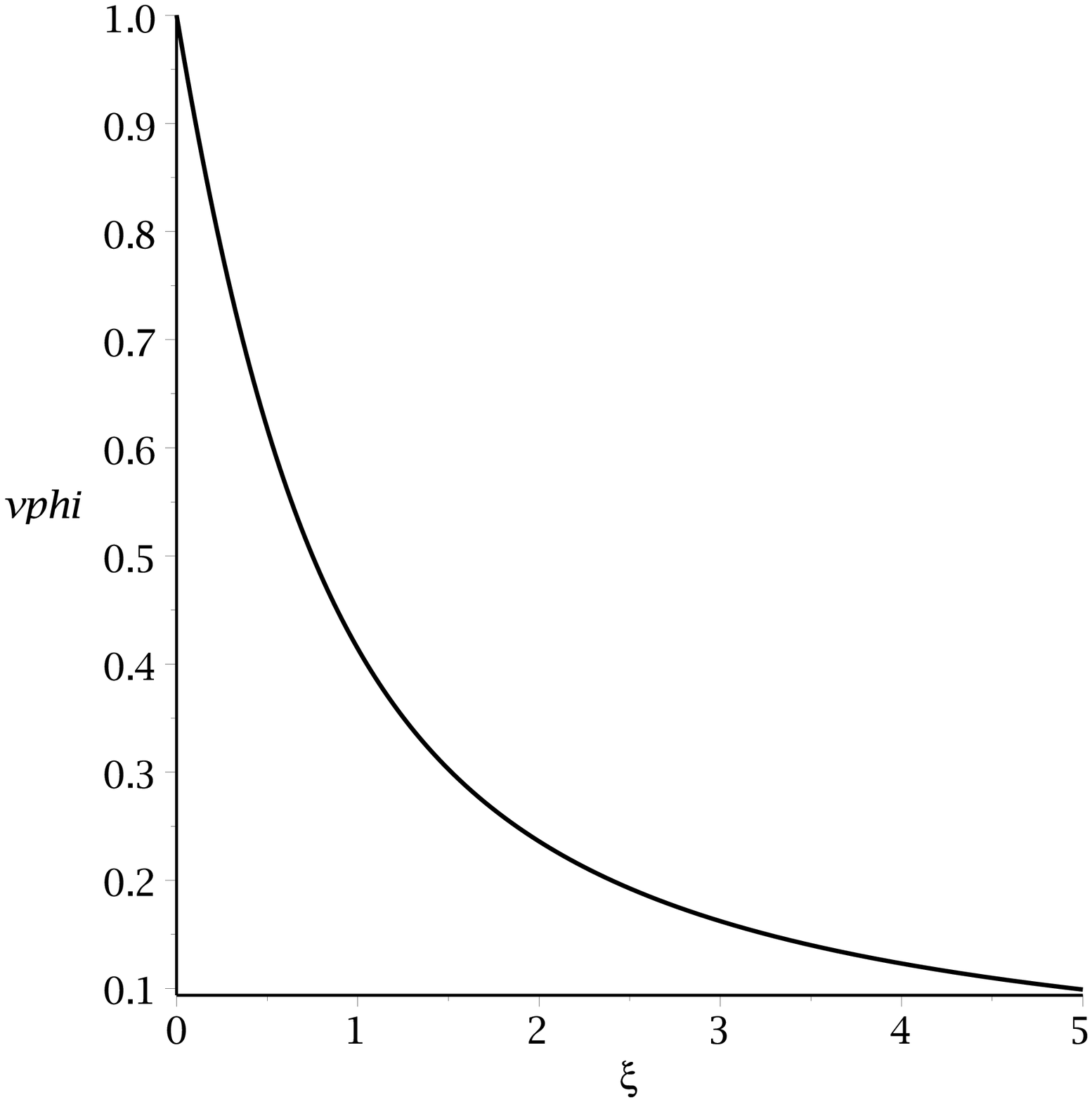}}}&
\rotatebox{0}{\scalebox{.4} 
{\includegraphics{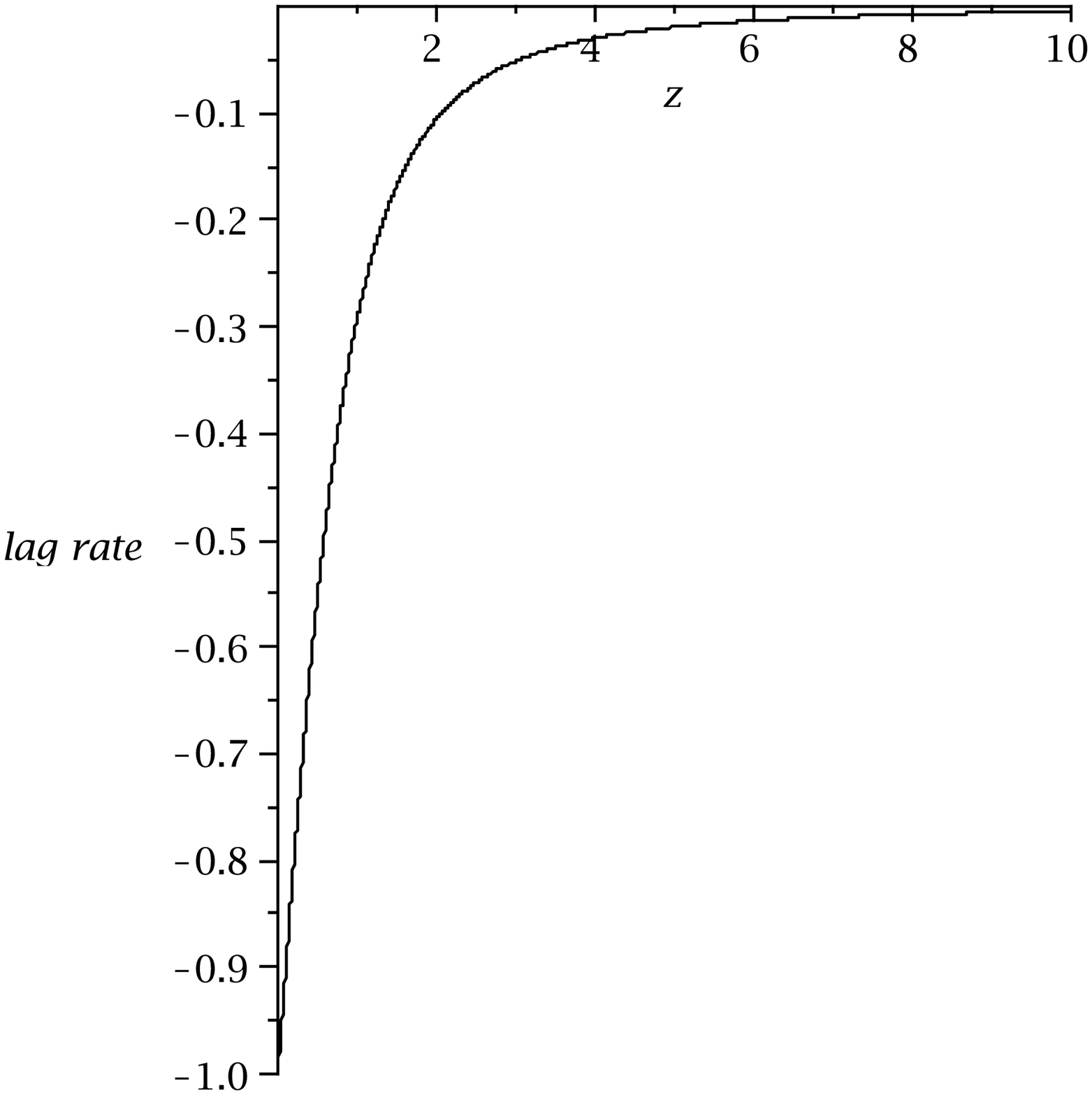}}}\\
{\rotatebox{0}{\scalebox{.4} 
{\includegraphics{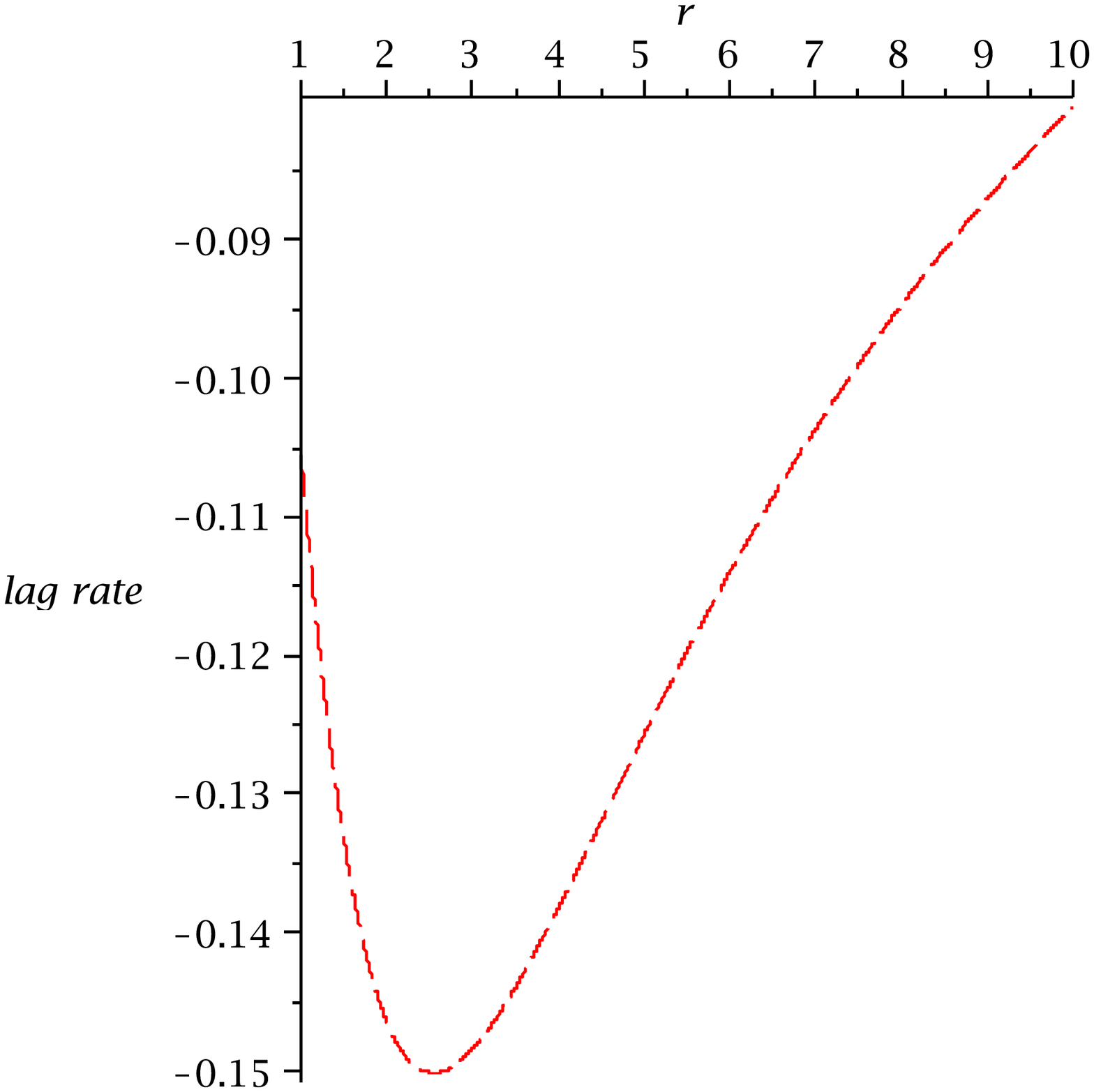}}}}
\end{tabular}
\caption{The figure shows at upper left the rotational velocity in the gaseous halo as s function of $\xi=z/r$ in units of $V$. As $r$ increases $z\propto r$ for the same fractional decrease in velocity. Hence the lag rate declines with radius. At fixed $r$ the curve shows the decline with $z$. At fixed $z$ the curve gives the dependence on $r$ when read from right to left. At upper right the actual lag rate is shown as a function of $z$ when $r=1$. Both scales are in Units of a fiducial scale $1/\delta$. The lag-rate is in Units of $\delta V$. At lower left the dashed curve shows the variation of the vertical lag-rate with radius at $z=2/\delta$. The Units are the same as at upper right.   }    
\label{fig:equipartlag}
\end{figure}

 The parameter free version of this curve is plotted in figure (\ref{fig:equipartlag}). One should remember that $\xi=z/r$ is constant on cones ($\theta=\pi/2-tan^{-1}(\xi)$ is the opening angle). The rotational velocity becomes small on a cone whose opening angle approaches $0^\circ$. The lag-rate (see equation (\ref{eq:vlag}) is also small there. At any $r$ this cone will be intercepted at sufficiently high $z$, while at any $z$ it will be intercepted at sufficiently small $r$. Such locations may well be unobservable in practice.

The vertical lag-rate is given from equation (\ref{eq:vphiexp}) by
\be
\partial_zv_\phi=\frac{V}{r}\left(\frac{z-\sqrt{r^2+z^2}}{\sqrt{r^2+z^2}}\right),\label{eq:vlag}
\ee
{\it which declines with increasing $r$ and with increasing $z$}. For example, at the disc $z=0$ so that $\partial_zv_\phi|_o=-V/r$. If $V=220$ km/sec and $r=8$ kpc, then the maximum lag rate at this radius is $\sim 27.5$ km/sec/kpc. In figure (\ref{fig:equipartlag}) we show this variation at $r=1/\delta$, where $1/\delta$ is a fiducial length scale. The lag-rate is in units of $\delta V$. We also show in the figure the variation with $r$ of the vertical lag-rate at a height $z=2/\delta$. There is a radius ($\sim 2.5/\delta$) at which the halo has a maximum vertical lag-rate, although this is not so evident at smaller $z$.

\section{Spatial Structure of the isotropic solution}\label{sect:fields}

\begin{figure}
\begin{tabular}{cc} 
&\rotatebox{0}{\scalebox{.8} 
{\includegraphics{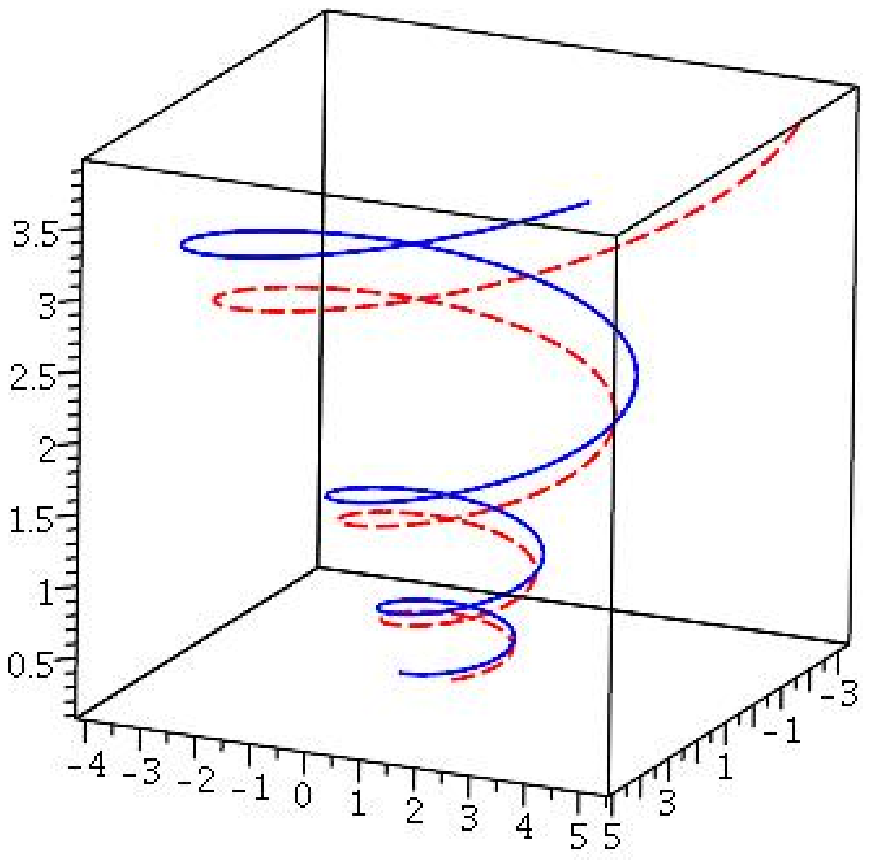}}}
\rotatebox{0}{\scalebox{.8} 
{\includegraphics{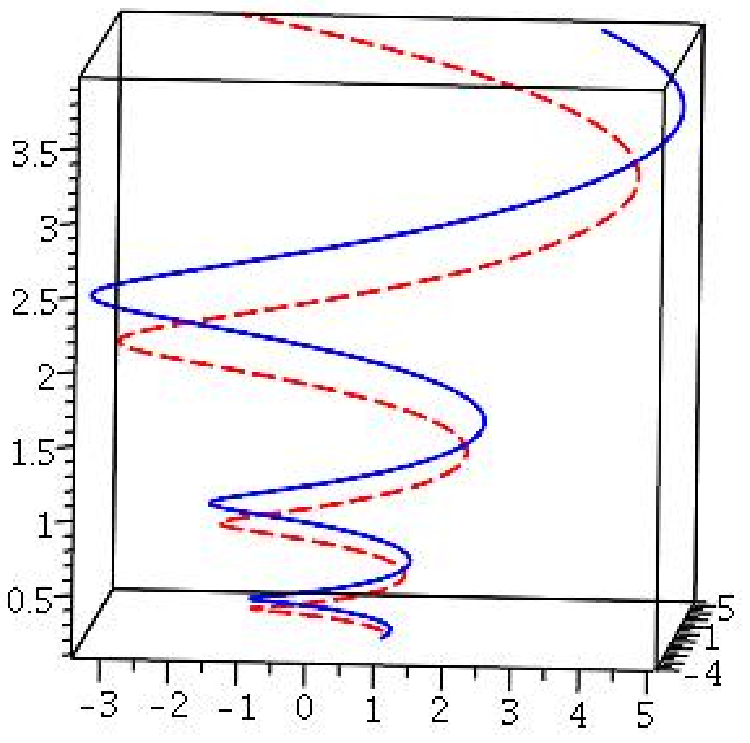}}}\\
&{\rotatebox{0}{\scalebox{.65} 
{\includegraphics{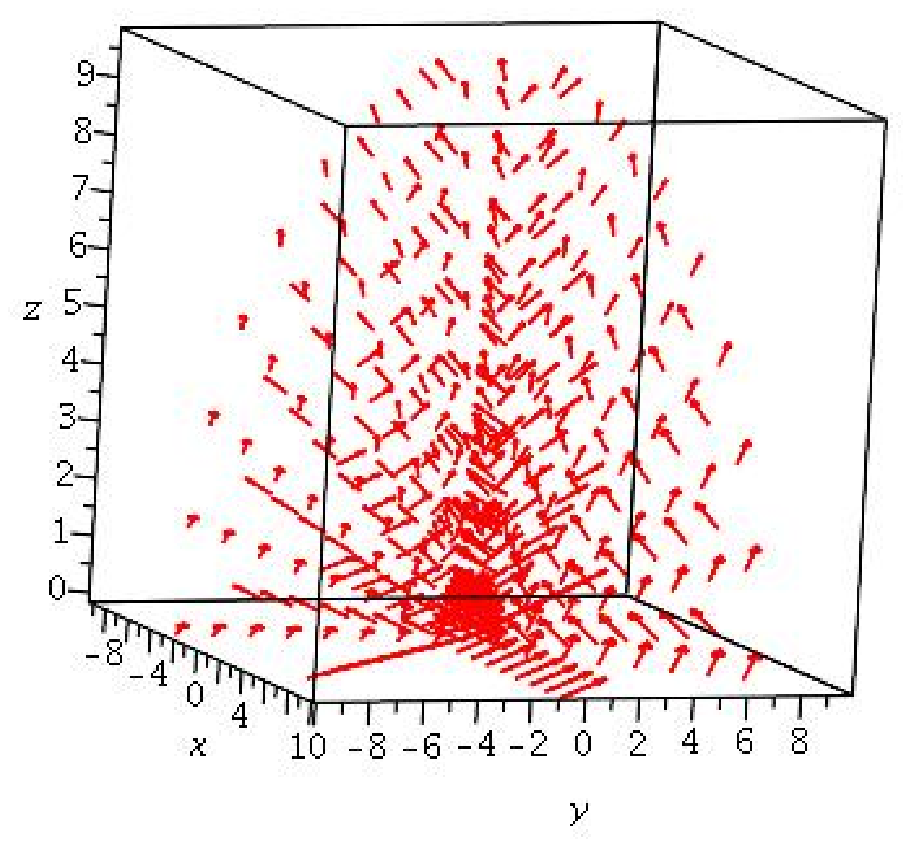}}}}
\rotatebox{0}{\scalebox{.65} 
{\includegraphics{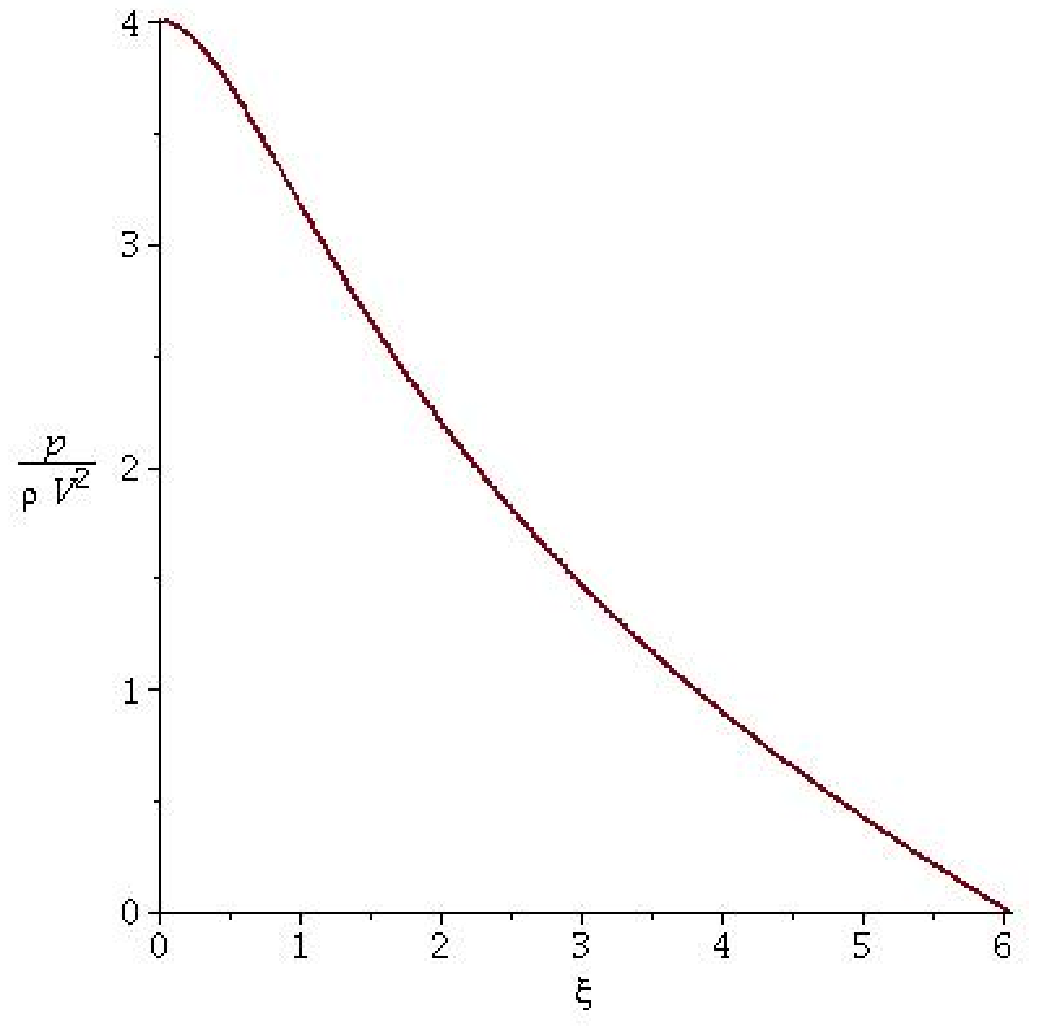}}}
\end{tabular}
\caption{ At upper left we show field lines that start in the plane at azimuths $\pm 0.5$ radians. The parameters are $c_3=0.1$ and $c_2=0.05$ in Units of the disc rotational velocity $V$. At upper right there is a slightly rotated view of the same field lines. At lower left we show a global velocity field line plot of the solution for the same $c_3$ but $c_2=0.2$. It is in the upper half plane with the polar angle $\le 0.25$ radians removed for clarity. The projected magnetic field (either parallel or anti-parallel to the velocity) shows an `X-type' field between the equatorial disc and the polar region. Faraday rotation due to the field along the line of sight will change sign across the axis. The pressure in Units of $\rho V^2$ is shown at lower right for the same parameters as in the upper figures plus $\lambda=1$, $\delta r=1$ and $p/(\rho V^2)=4$ at the equator. Since the radius is fixed in the plot, $\xi=z/r=z\delta$.   }    
\label{fig:equispace}
\end{figure}

In figure (\ref{fig:equispace}) we show some examples of the spatial structure of the velocity/magnetic field (we do not distinguish between parallel and antiparallel magnetic field). At upper left we show two field lines wrapping around the axis of the galaxy, starting at azimuth $\pm 0.5$ radians. The parameters $c_3=0.1$ and $c_2=0.05$, both in Units of $V$. The figure at upper right is a more nearly side-on image of the same system. The `zig-zag' projected shape of the field line is characteristic. One must imagine the field lines at different azimuths. 

The global field plot at lower left is  made for $c_3=0.1$ but $c_2=0.2$. The image is for the upper half plane only.  The angle extending to about $0.25$ radians from the axis has been cut out. We see the field parallel to the disc at the equator, subsequently rising to circulate on different conical structures. This configuration yields an `X-type' transverse field near the disc, which declines in $z$. Such a configuration  predicts that there should be a reversal in the sign of the Faraday rotation across the axis of the galaxy. This has been detected for the Milky Way \citep{Gfarr2015}. 

The pressure in Units of $\rho V^2$ is shown at lower right for the same parameters as for the field/stream lines, but $\lambda=1=\delta r$ and $(p/\rho V^2)\bigg|_o=4$. The pressure becomes negative under these conditions at $\xi=6.05$ (as per equation \ref{eq:xicrit}), that is $z=6.05/\delta$ where $\delta r=1$. Beyond this point the required equilibrium can not be maintained. If $1/\delta=2~kpc$, then this may well continue into the inter-galactic medium before becoming time dependent. The time dependence is likely the manifestation of Alfv\'en waves carrying the necessary flux of angular momentum before the steady state is reached.

The required temperature  of the halo gas just above the equatorial disc is $\sim 1-2~keV$. This may be reduced if the halo at this radius is not of equal gravitational importance to the disc ($\lambda <1$).  To a lesser extent this value can also be reduced by reducing the radial velocity at the disc, but this would also minimize the `X structure' of the fields. 

We have not used an equation of state in our discussions. This is normal for incompressible gas flows. However the pressure inferred is `dynamic' and need not be due only to hot gas. In particular, there may be a contribution from inhomogeneous sub-scale turbulence. However to make a significant contribution to the pressure, the turbulent velocities would need to be comparable to the mean flow. A constant pressure produced externally is possible, but in a realistic case such as that provided by 'ram pressure' axial symmetry would be broken at large scales.

\section{Real Galaxies}\label{sect:test}

In this section, we wish to compare our results, specifically, Eqns.~\ref{eq:vphiexp} and \ref{eq:vlag}, with measurements of real galaxies.   Such measurements are observationally challenging, given the faintness of halos in comparison to disks and also the spatial resolution that is required to resolve such halos in the $z$ direction. 
Nevertheless a growing number of galaxies are showing such effects (e.g. see references in the {\it Introduction}).

\begin{figure}
\begin{tabular}{cc} 
\rotatebox{0}{\scalebox{.5} 
{\includegraphics{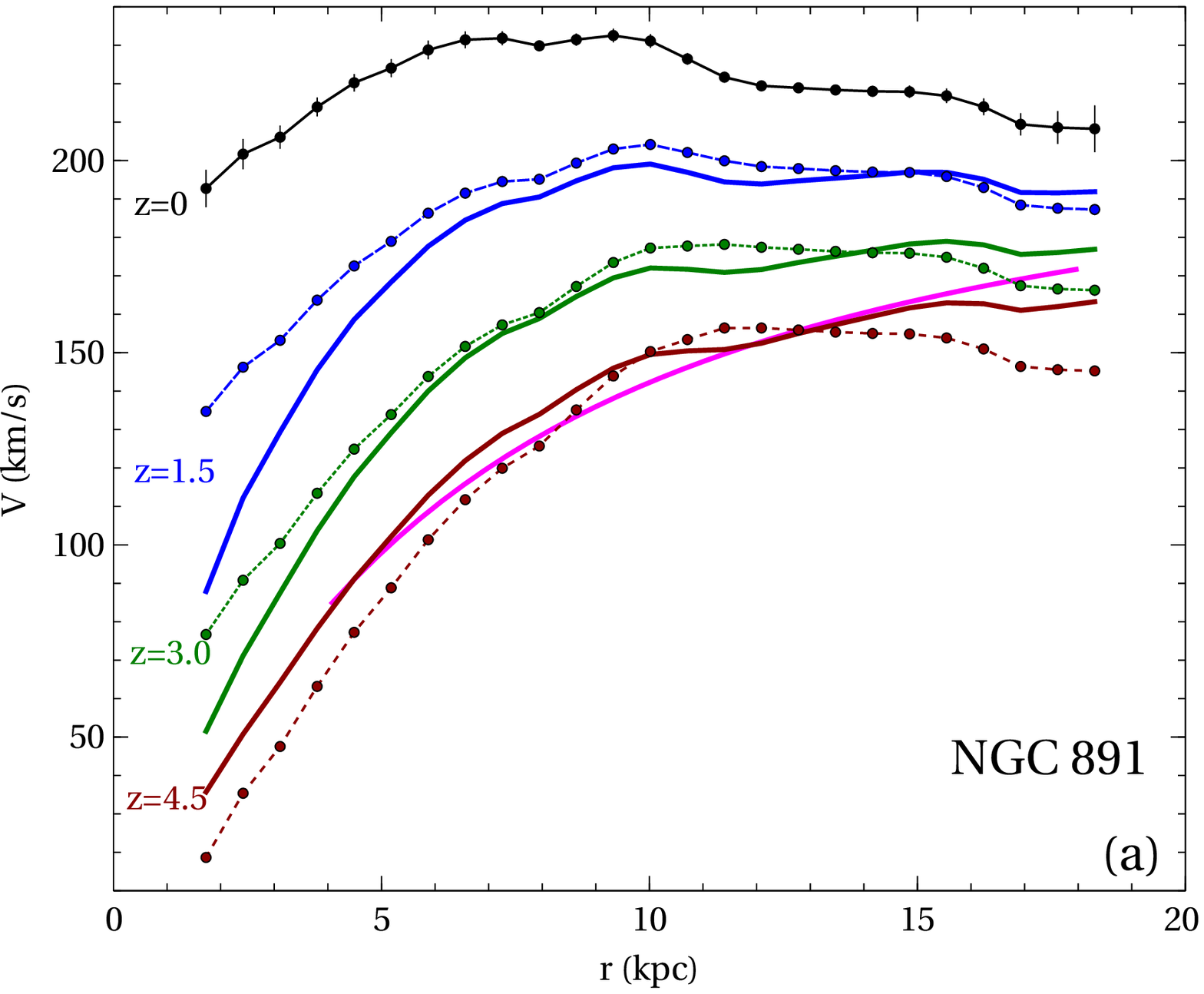}}}&
\rotatebox{0}{\scalebox{.5} 
{\includegraphics{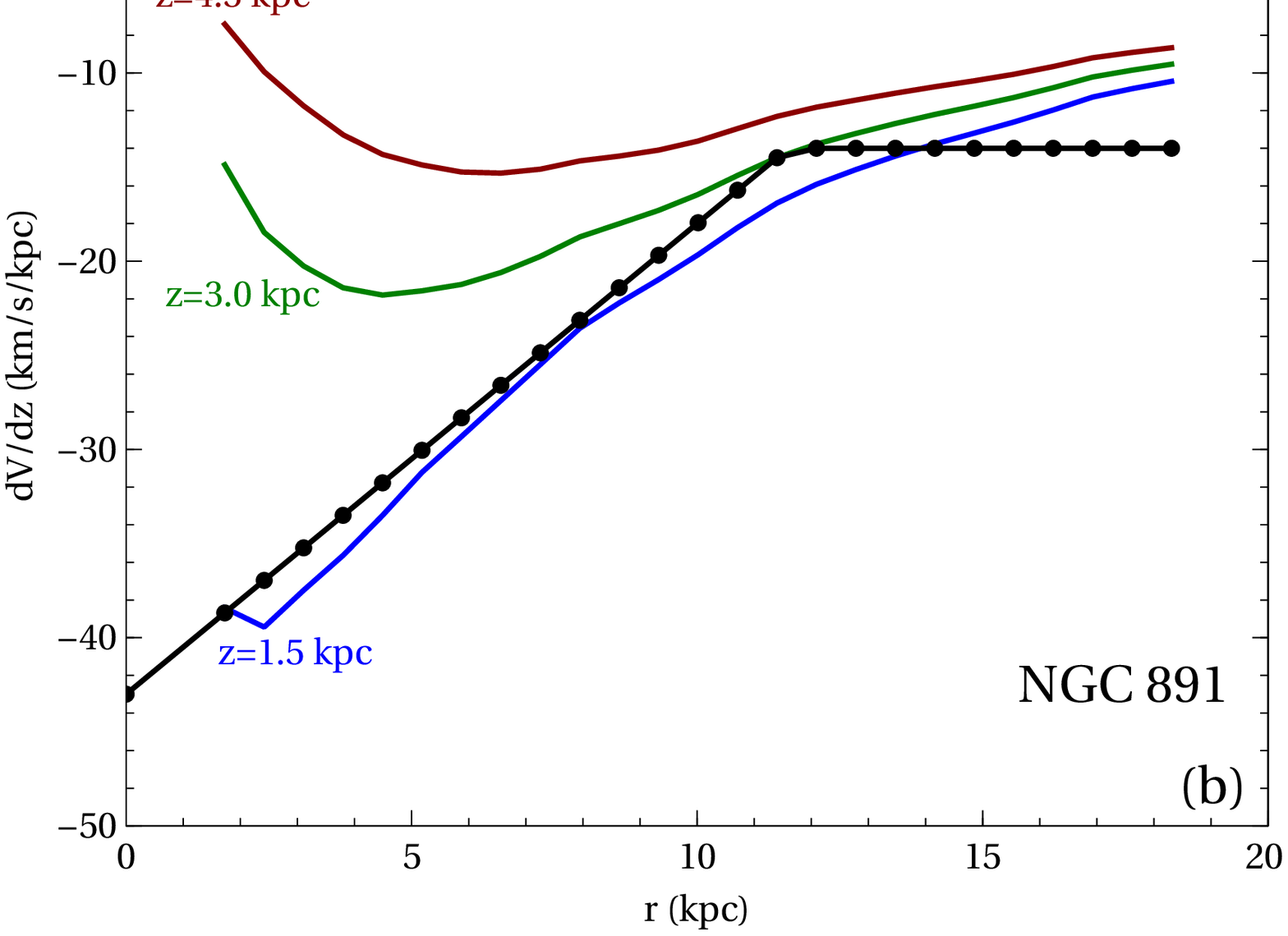}}}\\
\end{tabular}
\caption{{\bf (a)} Azimuthal velocity as a function of radius for the midplane (black curve) as well as three 
different heights above the midplane ($z=$ 1.5 kpc: blue, $z=$ 3.0 kpc: green, and $z=$ 4.5 kpc: red).  
The data are shown with curves and points (see Oosterloo et al. 2007) and our results, from Eqn.~\ref{eq:vphiexp}, 
are shown as solid curves without points in matching colours. The extra solid curve, shown in magenta, 
represents the $z$=4.5 kpc curve for which the rotation curve is taken to be completely flat with $V$=220 km/s.
{\bf (b)} Lag as a function of radius (black curve with points from Oosterloo et al. 2007).  The three curves without
points, from Eqn.~\ref{eq:vlag}, are shown for three different values of $z$  with colour coding as above.
}    
\label{fig:vphi}
\end{figure}

In Fig.~\ref{fig:vphi}a, we show $V_\phi$ as a function of radius for NGC~891.  The data, kindly supplied by Dr. Filippo Fraternali \citep[][see their Fig. 15]{oos07}, provided the best model fit to the HI observations in this galaxy.  They are shown as the curves with points for the midplane and for three heights, $z$, above the major axis. The solid curves (blue, green and red) show our values using Eqn.~\ref{eq:vphiexp}, where $V$ is the {\it underlying} measured in-disk rotational velocity (curve at $z=0$).  Since our coherent model assumes a constant midplane rotational velocity, we show an additional curve at $z$=4.5 kpc over the region from 4 to 18 kpc which corresponds to a constant midplane value of $V$=220 km/s (the average of the measured points).  These curves are reasonable representations of the data, given the simplicity of our  model.  Moreover the data were  extracted after modeling the observations by various empirical velocity components \citep{oos07}, of which ours was not one.

One should emphasize here that our model applies to the pressure dominant component of the halo gas, which may be a very hot component (see the discussion concerning the criterion \ref{eq:xicrit}). This is not seen directly in the HI observations employed here. In fact in the event that a hot halo gas does dominate, then  this paper only describes a magnetically mediated  lag of that component. Its transmission to the HI gas (perhaps clouds) would then have to be by way of the coupling mechanism explored in the series of papers \cite{FB2008},\cite{M2010},\cite{M2011} and references therein. A synthesis between this work and those must be left to future work. 
  
Lag {\it rates} as a function of radius for various galaxies have been nicely summarized in 
 Fig.~11 of \citet{ZR15}.  They show a wide range of values and tend to be crude, given measurement limitations, but the trend to lower lag rates with increasing $r$ is quite clear. The most extreme example belongs to NGC~4565, which shows a lag of -40 km/s/kpc at $r$=0.5$R_{25}$ to 0 km/s/kpc by $r$=$R_{25}$.  In NGC~891, as illustrated by the black curve with points in Fig.~\ref{fig:vphi}b, the gradient is steep (-43 km/s/kpc) at small radii, and decreases to -14 km/s/kpc at $r\,\approx\,$12 kpc
after which it is approximately constant with radius.  For NGC~891, as with other galaxies, only a single lag rate can be measured at any given radius so no measurements of the variation of the lag rate with $z$ are available.

Superimposed on Fig. ~\ref{fig:vphi}b are three curves depicting the
 lag rate for three different values of $z$, as calculated from Eqn.~\ref{eq:vlag}.  The $z$=1.5 kpc curve matches quite well but at higher value of $z$, there are significant departures at low $r$.  If existing lag rate data are more applicable to lower values of $z$, as might be expected from measurements which are weighted towards brighter emission, then the agreement is very good.  As data improve in the future, so that variations in lag rate with $z$ become available, then such curves may provide a way of testing this simple model.

\begin{figure}
\begin{tabular}{cc} 
\rotatebox{0}{\scalebox{.45} 
{\includegraphics{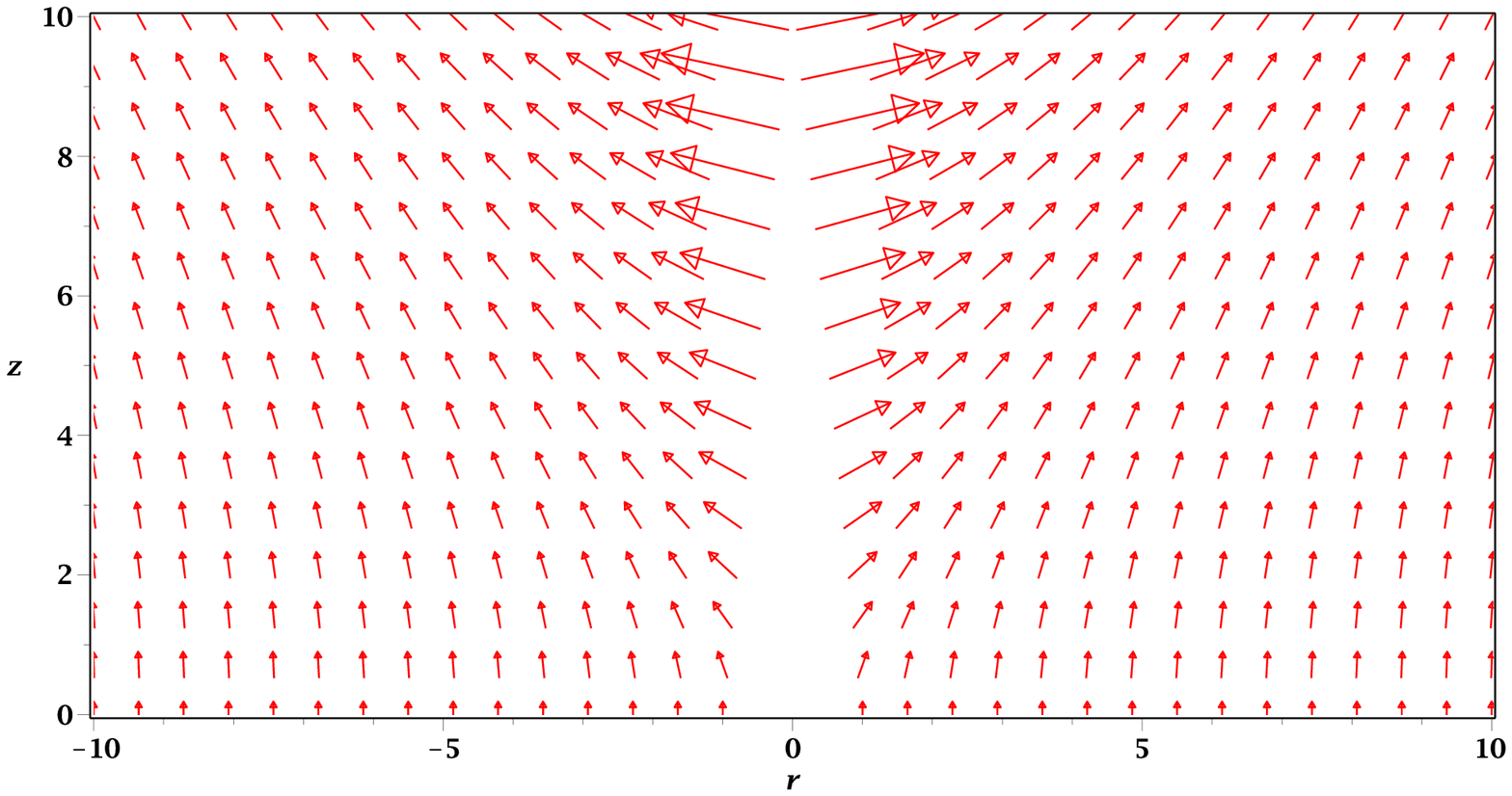}}}&
\rotatebox{0}{\scalebox{.3} 
{\includegraphics{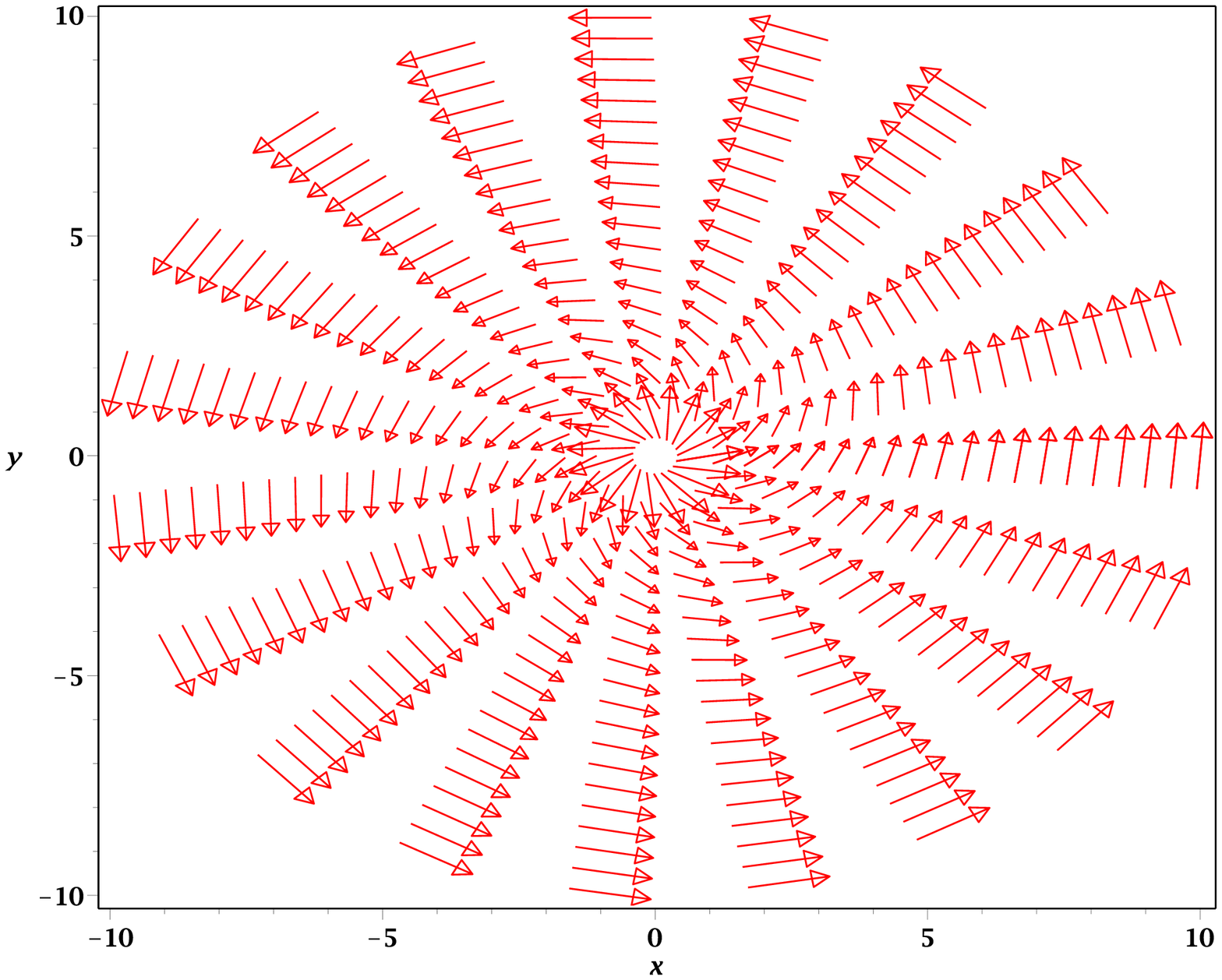}}}\\
\end{tabular}
\caption{{\bf (a)} A vector field in the {\it r-z} plane showing the X-type structure of either the magnetic
  field lines, B, or velocity, v, in the halo. The model is scale-free, but in this view, $r$ and $z$ can be
  thought of as being in units of kpc.  A single plane
  parallel to the sky is shown for $c_3=0.1$ and $c_2=0.2$ so that $v_z/v_r=2$. 
  {\bf (b)} A view of the same field looking from above (i.e. observing a plane parallel to the galaxy disk) for
  a height of 5 kpc.
}    
\label{fig:fields}
\end{figure}

 Finally, we show in
 Fig.~\ref{fig:fields}, the field structure derived from Fig.~\ref{fig:equispace} (lower left), shown here in simple single plane-views.  Since the B field and velocity field are in equipartition in the halo, the vectors represent either {\it B} or {\it v}.
 
 The left image shows the X-type field structure that results when observing the galaxy edge-on.  This image
 shows a single slice parallel to the sky plane.  In reality, to obtain an exact comparison between a
 real galaxy and our model requires an integration of all vectors along the line of sight weighted by emissivity.  This is beyond the scope of our model but the structure clearly shows the X-type behaviour that is observed in many edge-on systems. We note that the field vectors are vertical at the disc, then pass through an `X-type' region. Ultimately, as $\xi$ continues to increase vertically, the field will be dominated by the radial component and become parallel to the plane. The intermediate region is roughly where $\xi\sim c_2/c_3$, that is $\xi=2$ for the case shown. 

 On the right of figure \ref{fig:fields} is a `top-down' view of the halo velocity/magnetic field. The figure is really an axi-symmetric projection of the left hand figure onto  a plane parallel to the equatorial plane  at a vertical height of $5$ kpc. Assuming that one observes close to this plane as is the case for edge-on galaxies, the figure illustrates how the direction of the magnetic field would reverse depending on which side of the galaxy center is being observed.  Such a reversal is a prediction of this model and should be observable in Faraday Rotation. The figure also illustrates the initial decrease with radius of the circular velocity, and the subsequent increase to form the Taylor-Proudman column at large radii.

\section{Discussion and Conclusions}\label{sect:conclusions}

Our basic conclusion is that the halo lag phenomenon may indicate a magnetic coupling between the halo gas and the inter-galactic medium. The evidence is summarized in figures (\ref{fig:vphi}) and (\ref{fig:fields}a). Ultimately the coupling is between the disc of the galaxy and the distant environment, because we require the magnetic field to be `anchored' in both locations. This explanation requires  a large scale magnetic field, such as might be produced by dynamo action during the formation of the galaxy. Such dynamo action has not been included here, however.

We have explored this possibility using a simple, scale-invariant, analytic, model that assumes a steady state. Equipartition between the kinetic energy density of the halo gas and the magnetic energy density there is also part of the model. The gravitational fields of a Mestel disc and of an isothermal collisionless halo are included. The disc has a strictly constant rotational velocity $V$ and all gravitational, magnetic, and velocity fields incorporate this constant. Hence the model can not be applied to central regions of the galaxy. This also implies a height limit in $z$, because ultimately an unreasonable radial velocity associated with the central region of the galaxy is encountered. Fortunately this is normally beyond the observable region.

The scale-invariance requires the viscosity of the constant density halo gas to be proportional to the radius. A consequence of this assumption together with that of constant density is the development at large radius of a Taylor-Proudman column. This is uniform in $z$ and rotates with the speed $V$. It is likely that the assumptions break down before this column is fully established, but it does affect our predicted kinetic behaviour (see figures (\ref{fig:vphi}) and (\ref{fig:fields}).

The equipartition assumption allows us to infer the required magnetic field strength from equation (\ref{eq:magfield}), given the density and velocity of the relevant component. As estimated in the introduction, the field at the disc would need be as high as $10$ $\mu G$ if a gas of total gas density $0.01$ $cm^{-3}$ is rotating at the full disc velocity. This is rather higher than observed in most galaxies \cite{Beck2016}, but we have seen that it is rather less ($\sim 5$ $\mu G$) for a more rarefied hot component. In any case the main reason for the equipartition assumption is analytic simplicity. There is no reason to believe that a magnetic field less than the equipartition value would not have a similar effect over a longer time. In fact in view of the discussion in the next paragraph, a weaker magnetic field may even be more favourable to the basic mechanism.

The steady state assumption removes the causal path to equilibrium. The required coupling and energy equipartition may originate in the formation stage of the galaxy\\
 \citep{PMS2014}. However for our mechanism to be plausible, the drag on the gaseous galactic disc should {\it not} succeed in destabilizing it  
for a long period extending backwards from the present epoch. This can require the time of magnetic disc-environment coupling to occur at an epoch later than that of  actual galaxy formation. A crude estimate for a milky-way type galaxy suggests that this may be as late as red-shift $\sim 1$, depending principally on the mass of the gaseous disc and the strength of the magnetic stress. The simulation in \citet{PMS2014} concludes that the epoch of developed galactic magnetic field is at redshift $\le\sim 2$, but we would probably need the  coupling invoked here to arise later than this.     

 Intuitively the coupling development admits two forms of the equipartition assumption, which we refer to as `isotropic' and `anisotropic'. The first form intuitively conforms best to coupling involving lagging accreting gas in the halo,because  descending magnetic and velocity fields are naturally held behind the disc rotation by the inter-galactic drag. The second form with $B_\phi$ parallel to $v_\phi$ while $B_z$ is anti-parallel to $v_z$ establishes the same intuitive coupling for lagging gas ejected from the disc.

The behaviour vertically `across' (e.g. north/south) the disc depends on whether the global magnetic field is mainly dipolar or perhaps more nearly quadrupolar. However for a largely dipolar field, the two forms of equipartition apply in the same manner. That is the isotropic case is intuitively appropriate for lagging gas descending to the disc, while the anisotropic case describes intuitively lagging ejected gas.

The principal results that we have tested are those of equations (\ref{eq:vphiexp}) and (\ref{eq:vlag}). These are displayed against the data for NGC891 in figure (\ref{fig:vphi}).There are some slight deviations only at small $r$, using the underlying rotation curve for $V(r)$  We have also indicated the  prediction at a height of $4.5$ kpc if the rotation curve is flat. The agreement seems satisfactory except at large disc radii. This might be expected from our remarks regarding model limitations previously in this section. 

On the right of the figure we show the actual lag rate from equation (\ref{eq:vlag}). The observed lag is one point at each radius that we have fitted with our formula at a height of $1.5$ kpc. The expected deviations at large and small radii become much more pronounced at larger heights. Observations with improved resolution may detect these deviations.
 
Our final figure (\ref{fig:fields}) emphasizes on the left the `X-type' magnetic field (and velocity field) configuration. We show simply an $r-z$ plane without taking into account an integration along the line of sight. This is left for more detailed modeling elsewhere.  The angle of the field lines in the figure depends on the ratio of the vertical to radial velocities. We see that the X configuration is predicted to be an intermediate phenomenon, lying between vertical fields near the disc and more radial fields at high $z$. 

Figure (\ref{fig:equispace}) displays the field structure in three dimensions, as well as a typical halo pressure curve that is required to maintain the equilibrium. The upper figures require adding field lines in azimuth to form a complete picture. The 3d field is illustrated at lower left, but is more difficult to interpret. The important prediction is that the sign of the Faraday rotation should reverse across the axis of the galaxy.



\section*{Acknowledgements}

We are grateful to Fillipo Fraternali for making available the data in Fig.~\ref{fig:vphi}.
Also, many thanks to R. Beck, R. Walterbos, Q. D. Wang, and M. Krause
for helpful comments.
JI would like to thank the Natural Sciences and Engineering Research Council of Canada for a Discovery Grant.









\bsp	
\label{lastpage}
\end{document}